\documentclass[twocolumn,pra,aps,showpacs]{revtex4}
\usepackage{amsmath}
\usepackage{color}
\usepackage{appendix}
\usepackage{graphicx,bm,dsfont}
\usepackage{CJK} 

\begin{document}

\begin{CJK*}{GBK}{song} 

\title{Tunable photon scattering by an atom dimer coupled to a band edge of a photonic crystal waveguide}

\author{Guo-Zhu Song,$^{1,}$\footnote{Contact author: songguozhu@tjnu.edu.cn} Lin-Xiong Wang,$^{1}$ Jing-Xue Zhang,$^{2}$ and Hai-Rui Wei$^{3}$ }

\address{$^{1}$College of Physics and Materials Science, Tianjin Normal University, Tianjin 300387, China\\
$^{2}$School of Public Health, Tianjin University of Traditional Chinese Medicine, Tianjin 301617, China\\
$^{3}$School of Mathematics and Physics, University of Science and Technology Beijing, Beijing 100083, China
}
\date{\today }

\begin{abstract}
Quantum emitters trapped near photonic crystal waveguides have recently emerged as an
exciting platform for realizing novel quantum matter-light interfaces. Here we study tunable photon scattering
in a photonic crystal waveguide coupled to an atom dimer with an arbitrary
spatial separation. In the weak-excitation regime, we give the energy levels and their
decay rates into the waveguide modes in the dressed basis, which both depend on the distance
between the two atoms. We focus on the Bragg case and anti-Bragg case, where
subradiant and superradiant states are produced and perfect transmission with a $\pi$ phase
shift may occur on resonance. We observe quantum beats in the photon-photon correlation function
of the reflected field in the anti-Bragg case. Moreover, the frequencies of quantum beats can
be controlled due to the tunability of the bound states via the dispersion engineering of the
structure. We also observe directional photon emission in the anti-Bragg case and give the dynamic mechanism of the perfect transmission.
We quantify the effects of the system imperfections, including the deviation in
the distance between the two atoms and the asymmetry in the atomic decay rates into the waveguide
modes. With recent experimental advances in the superconducting microwave transmission lines,
our results should soon be realizable.
\end{abstract}

\maketitle

\section{Introduction}

The interaction between light and atoms is at the heart of modern quantum optics \cite{PLodahl2015rmp}.
Strong light-atom interaction has been realized in cavity quantum electrodynamics (QED) \cite{Haroche2006,2006Walther,Reise2015}.
Of the different platforms, waveguide QED is a particularly promising candidate for exploring light-matter interaction and quantum optical phenomena \cite{Shen2007PRL,Diby2017rmp,DEChang2018,Shere2023Rev}.
In the framework of waveguide QED, quantum emitters can couple to, and interact via, the continuous bosonic modes in one-dimensional (1D) waveguides.
The waveguide QED platform has been realized in a variety of systems, such as atoms and quantum dots coupled to optical nanofibers \cite{DayanScience2008,Vetsch2010prl,RausPRL2011,DReitz2013PRL,Petersen2014,Liao2015pra,Liao2016,HLsorensen2016,Cheng2017pra,song2017pra,PSolano2017,Litao2018,song2021OE,Liedl2023prl}, photonic crystal waveguides (PCWs) \cite{TLHansen2008,AGoban2015PRL,Chang2011njp,MArcari2014prl,TudelaNAT2015,Yu2019pnas,Burgersa2019}, coupled resonator waveguides \cite{Lzhou2008,Liao2010pra,Rabl2016pra,wang2020PRL,Yang2020OE,Wang2023PRA}, and surface plasmon nanowire \cite{Chang2007nap,AkimovNature2007,Tudela2011prl,Akselrod2014,Anj2024PRA}, as well as superconducting artificial atoms coupled to microwave transmission lines \cite{WallraffNature2004,ShenPRL2005,AstafievScience2010,LooSci2013,GuPR2017,Das2017PRL,Kockum2018,Liaoyin2022pra,Liao2023pra,Joshi2023prx,Kannan2023NP,Jingj2024NJP}. In the many-body regime of waveguide QED, collective excitation states, such as subradiant and superradiant states, possess surprising physical properties and have attracted much attention \cite{klalum2013,zhang2019prl,Henriet2019pra,Ke2019prl,Albrecht2019njp,zhang2020prr,Nie2020prl,Holzin2022prl,Tiranov2023sci,Nie2023PRL,Nie2024ctp,Lu2024qst,Lu2024PRA}.

Due to tailored dispersion relations, PCWs with periodic dielectric structures provide a
unique avenue to control and modify light propagation \cite{SJohn1990prl,SJohn1990prb,SJohn1995prl}.
Near a band edge, the guided modes in PCWs travel slowly and the atom-photon coupling can be strongly
enhanced. In particular, when the atomic transition lies in a band gap, localized atom-photon bound states
appear around the atoms and result in tunable long-range atom-atom interactions \cite{Hung2013,JSDouglas2015,ShiT2016,jsDoug2016prx,Shah2016Op,Shi2018njp,Kumar2023}.
The localization of the bound state is determined mainly by the properties of the band edge, and the frequency detuning
between the qubit and the band edge. These interesting features have led to qualitatively new physics and have attracted much
attention for applications in quantum optics and many-body physics \cite{Song2018,Bello2019,Song2019,Kim2021prx,Bello2022prx,Wang2024res}.
In experiment, Liu and Houck observed the bound state in a system in which a superconducting transmon qubit is
coupled to a stepped-impedance microwave photonic crystal \cite{YLiu2017}, and then Sundaresan \emph{et al.} realized tunable long-range interactions between two transmon qubits in a similar system \cite{NMSun2019}. Later, Scigliuzzo \emph{et al.} observed coherent population exchange and cross-Kerr interaction between two atom-photon bound states by embedding two frequency-tunable artificial atoms in an array of Josephson-junction resonators \cite{Scigliu2022}. Recently, Zhang \emph{et al.} realized an extended version of the Bose-Hubbard model with tunable hopping range based on a lattice of superconducting transmon qubits coupled to a photonic-band gap metamaterial \cite{zhang2023sci}.
These recent experimental advances in PCW systems pave a tunable way for quantum simulation of many-body physics \cite{JSDouglas2015}. However, for quantum emitters coupled to a PCW, the role of collective excitation states in the enhancement of photon scattering and quantum nonlinear processes remains unclear.

In this work, we study photon scattering of an atom dimer coupled to a band edge of a PCW.
Our system is the minimal model that seeds long-range coherent interactions between atoms
and allows their quantum phenomena to be probed. In the weak-excitation regime, by diagonalizing
the effective Hamiltonian, we give the energy levels and their decay rates into the waveguide
modes in the dressed basis, which depend on the distance between the two atoms. Moreover, the
energy levels of the two dressed states are controllable due to the tunable interactions between atoms arising from the bound states.
In particular, we focus on the Bragg case and anti-Bragg case, where some novel quantum phenomena appear in
the atom-waveguide system. For the Bragg case, with free-space dissipation of atoms, one dressed
state is a superradiant state, and the other becomes a dark state. However, for the anti-Bragg case, perfect transmission
with a $\pi$ phase shift occurs on resonance under some specific conditions, which is different from the single-qubit case
\cite{ShenOL2005,Chang2007nap} and two-qubit case with no bound states \cite{zheng2013PRL}.
Furthermore, we give the photon-photon correlation function of the reflected field for the Bragg and anti-Bragg cases.
The results show that, for the Bragg case, the photon-photon correlation function in our system is similar to that for a single qubit with no bound
state \cite{Chang2007nap,Zheng2010PRA}. This is because, in this case, only the superradiant state is coupled to the waveguide modes.
However, for the anti-Bragg case, quantum beats (oscillations) between antibunching and bunching occur in the photon-photon
correlation, and the frequency of the oscillation can be controlled due to the tunability of the bound states.
Also, we observe directional photon emission in the anti-Bragg case in our system and reveal the dynamic mechanism of the perfect transmission.
Additionally, we discuss the effects of the imperfections in realistic atom-waveguide systems, including the deviation in atomic spacing and the asymmetry in the decay rates of atoms into the waveguide modes.

This paper is structured as follows. First, in Sec. \ref{model}, we review the effective Hamiltonian
governing atoms coupled to a PCW and rewrite the Hamiltonian with two dressed-state operators.
Second, in Sec. \ref{body}, in the weak-excitation limit, we present the energy shifts and effective decay rates of our system with an arbitrary atomic spatial separation.
In particular, we study the quantum phenomena of our two-atom model in the Bragg and anti-Bragg cases, and analyze the effects of the imperfections in realistic
atom-waveguide systems. Finally, we discuss the feasibility of our system with current experiments and summarize the results in Sec. \ref{discussion}.

\section{Model and Hamiltonian} \label{model}

\begin{figure}
\centering\includegraphics[width=7.0cm]{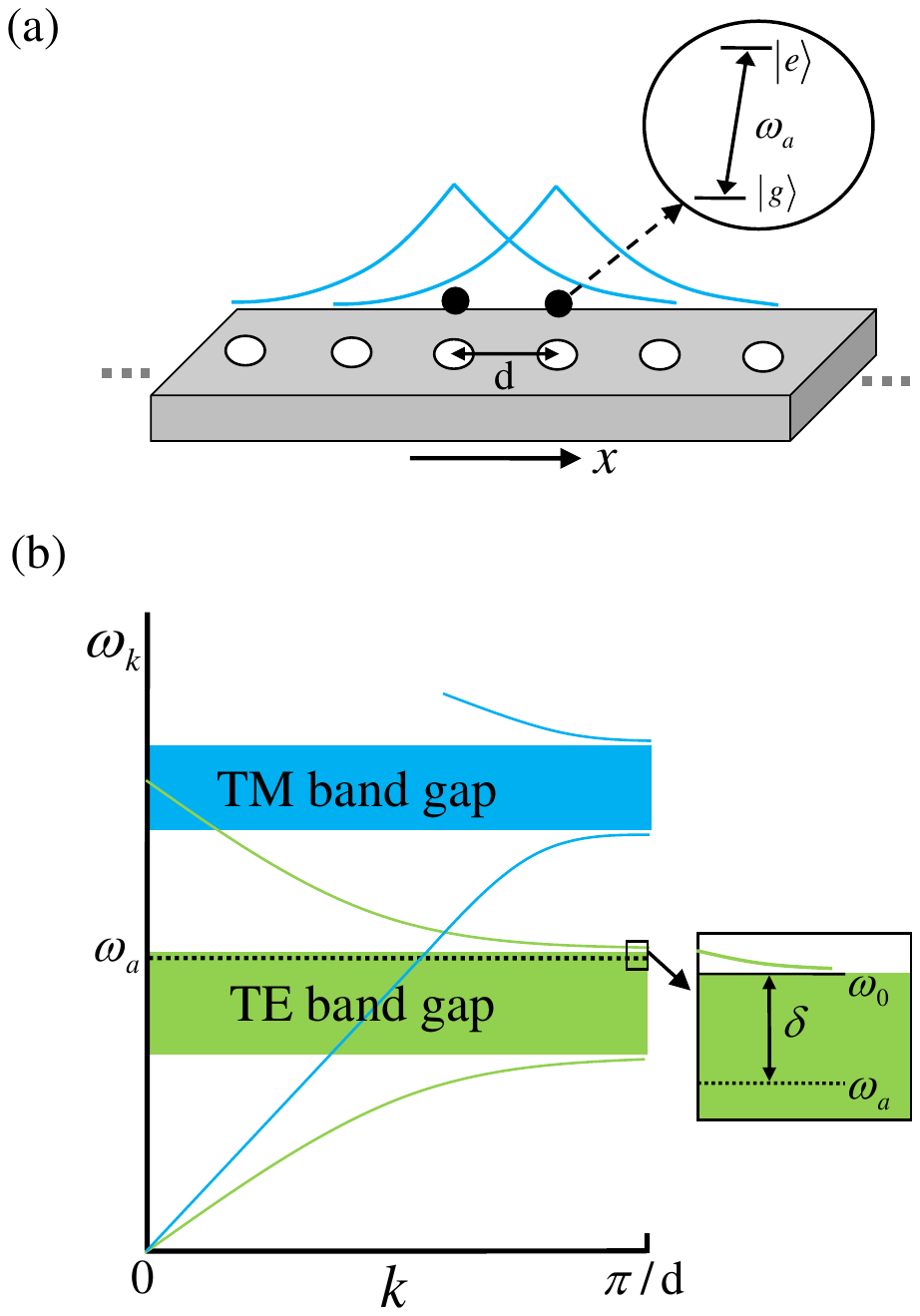}
\caption{ (a) Schematic diagram for an atom dimer (black dots) coupled to a PCW with unit cell length $d$.
A photon-atom bound state exists when $\omega_{a}$ is in a band gap, whose photonic component (blue envelope) is spatially centered at the atom.
(b) Band structure of the TM mode (blue) and TE mode (green) in a 1D PCW, showing the relation between the guided-mode frequency $\omega_{k}$ and Bloch wave vector $k$. The atomic frequency $\omega_{a}$ is in the band gap of the TE mode (green region), and in a linear-dispersion region of the TM mode (blue solid line). $\omega_{a}$ is close to the upper band edge frequency $\omega_{0}$ (wave vector $k_{0}$) with the frequency separation $\delta=\omega_{0}-\omega_{a}$.  }\label{figure1}
\end{figure}

\begin{figure*}[tpb]  
\begin{center}
\includegraphics[width=16.0 cm,angle=0]{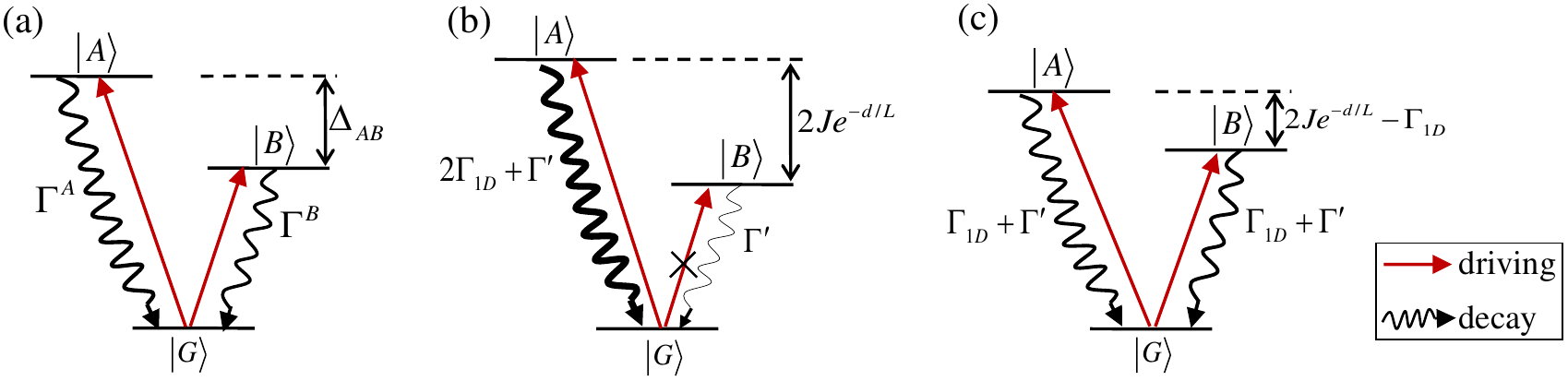}
\caption{ Energy levels and their decay rates into the waveguide modes in the dressed basis $\{$$|A\rangle$, $|B\rangle$$\}$ for (a) the general case, (b) Bragg case $k_{a}d=\pi$ and (c) anti-Bragg case $k_{a}d=0.5\pi$. $\Gamma^{A}=\Gamma_{_{1D}}+\Gamma'-\Gamma_{_{1D}}\cos(k_{a}d)$, $\Gamma^{B}=\Gamma_{_{1D}}+\Gamma'+\Gamma_{_{1D}}\cos(k_{a}d)$, and $\Delta_{_{AB}}=2Je^{-d/L}-\Gamma_{_{1D}}\sin(k_{a}d)$. The transition frequencies of states $|A\rangle$ and $|B\rangle$ are $\omega^{A}=\omega_{a}+J+Je^{-d/L}-\Gamma_{_{1D}}\sin(k_{a}d)/2$ and $\omega^{B}=\omega_{a}+J-Je^{-d/L}+\Gamma_{_{1D}}\sin(k_{a}d)/2$, respectively. $|G\rangle=|\text{g}\text{g}\rangle$ represents a state in which  the two atoms are both in the ground state $|\text{g}\rangle$, and $|ee\rangle$ is ignored for weak probe fields. } \label{figure2}
\end{center}
\end{figure*}

We model a pair of two-level atoms with ground and excited states $|\text{g}\rangle$ and $|e\rangle$ coupled to a 1D PCW, as shown in Fig. \ref{figure1}(a). A PCW is a periodic dielectric structure whose refractive index is modulated periodically \cite{JDJoan2008book}. This particular photonic structure can separate the modes into two different polarizations: the transverse-magnetic (TM) modes, whose polarization is perpendicular to the plane of the waveguide, and the transverse-electric (TE) modes, which are polarized in the plane of the waveguide. Both TM and TE modes are transverse to the direction of propagation, i.e., perpendicular to the $x$ axis. For some frequency windows, the modulation prevents the propagation of photon field input into the structure. This gives rise to the existence of band gaps in the dispersion relation for the photonic modes, as depicted schematically in Fig. \ref{figure1}(b).
Specifically, when an excited atom is trapped near the PCW and the resonance frequency $\omega_{a}$ lies in the TE band gap, the excited atom cannot radiate into the TE modes.
However, this effect can generate an exponentially localized atom-photon bound state with characteristic length $L$ around the excited atom, which facilitates interactions with proximal atoms via virtual photons with characteristic strength $J$ \cite{JSDouglas2015,jsDoug2016prx}. It has been proved that such photonic cloud acts as a real cavity mode, which enables long-range coherent interactions between the atoms. In addition, we assume $\omega_{a}$ is away from the TM band gap and thus the transition $|\text{g}\rangle \leftrightarrow |e\rangle$ is coupled to the TM modes.
The dynamics of the whole system including two atoms and the 1D waveguide modes can be described by an effective non-Hermitian Hamiltonian ($\hbar=1$) \cite{Chang2012,EWAN2017,Albrecht2017njp}
\begin{eqnarray}     \label{eqa1}       
H_{1}=&&\!\!\!\!\!-\!{{\sum\limits_{j=1}^2}}(\Delta+i\Gamma'/2)\sigma_{ee}^{j}\!-\!i\frac{\Gamma_{_{1D}}}{2}{{\sum\limits_{j,k=1}^2}}e^{ik_{a}|x_{_{j}}-x_{_{k}}|}\sigma_{e\text{g}}^{j}\sigma_{\text{g}e}^{k}\nonumber\\
&&\!\!\!\!\!+J{{\sum\limits_{j,k=1}^2}}(-1)^{\theta_{jk}}e^{-|x_{_{j}}-x_{_{k}}|/L}\sigma_{e\text{g}}^{j}\sigma_{\text{g}e}^{k}.
\end{eqnarray}
Here, $\Delta=\omega_{p}-\omega_{a}$ is the detuning between the frequency
$\omega_{p}$ of the  probe field (input from the left with wave vector $k_{p}$ and Rabi frequency $\Omega_{p}$) and the atomic resonance frequency $\omega_{a}$ with wave vector $k_{a}$. $\Gamma'$ is the decay rate of single atom into the free space, and the atomic operator $\sigma_{ab}^{j}=|a_{j}\rangle \langle b_{j}|$ with $a,b=\text{g},e$ and $j=1,2$. $\Gamma_{_{1D}}$ denotes the single-atom decay rate into the TM modes, and $x_{_{j}}$ represents the position of the atom $j$. Here, we assume the atoms are located at the antinodes of the TE mode, whose Bloch wave function is $E(x)\sim \cos(kx)$. This results in the phase factor $\theta_{jk}=(x_{_{j}}+x_{_{k}})/d$ and maximizes the interaction.
In this work, we probe the atomic transition by a weak coherent field input from the left ($\Omega_{p}\ll\Gamma_{_{1D}}$) and the corresponding driving Hamiltonian is $H_{2}=-\Omega_{p}{{\sum\limits_{j}^2}}(e^{ik_{p}x_{_{j}}}\sigma_{e\text{g}}^{j} + \mathrm{H.c.})$ \cite{CanevaNJP2015}. In principle, to preserve the norm, quantum jumps should be included to describe the atomic dynamics. However, we here mainly focus on the field response in the weak-excitation limit ($\Omega_{p}\ll\Gamma_{_{1D}}$), and quantum jumps can be neglected \cite{Albrecht2017njp}. Thus, the atomic dynamics are governed by the total Hamiltonian $H=H_{1}+H_{2}$.

In the following, we set $x_{1}=0$ and $x_{2}=d$ for simplicity. To give a more transparent form of the total Hamiltonian, we define two dressed-state operators \cite{klalum2013}
\begin{eqnarray}     \label{eqa2}       
\sigma_{e\text{g}}^{A}=\frac{-\sigma_{e\text{g}}^{1}+\sigma_{e\text{g}}^{2}}{\sqrt{2}},\;\;\;\;\;\sigma_{e\text{g}}^{B}=\frac{\sigma_{e\text{g}}^{1}+\sigma_{e\text{g}}^{2}}{\sqrt{2}}.
\end{eqnarray}
In the new basis set, the total Hamiltonian can be rewritten as $H=H^{A}+H^{B}$, with

\begin{eqnarray}            
H^{A}\!\!=\!\!\!\!\!\!&&\big\{\!-\!\Delta\!+\!J\!+\![Je^{-d/L}\!-\!\frac{\Gamma_{_{1D}}}{2}\sin(k_{a}d)]\!-\!i\frac{\Gamma^{A}}{2}\big\}\sigma_{e\text{g}}^{A}\sigma_{\text{g}e}^{A}\nonumber\\
&&\!\!-\frac{\Omega_{p}}{\sqrt{2}}[(e^{ik_{p}d}\!-\!1)\sigma_{e\text{g}}^{A}+\mathrm{H.c.}], \label{eqa3a} \\
H^{B}\!\!=\!\!\!\!\!\!&&\big\{\!-\!\Delta\!+\!J\!-\![Je^{-d/L}\!-\!\frac{\Gamma_{_{1D}}}{2}\sin(k_{a}d)]\!-\!i\frac{\Gamma^{B}}{2}\big\}\sigma_{e\text{g}}^{B}\sigma_{\text{g}e}^{B}\nonumber\\
&&\!\!-\frac{\Omega_{p}}{\sqrt{2}}[(e^{ik_{p}d}\!+\!1)\sigma_{e\text{g}}^{B}+\mathrm{H.c.}]. \label{eqa3b}
\end{eqnarray}
Here, $\Gamma^{A}=\Gamma_{_{1D}}+\Gamma'-\Gamma_{_{1D}}\cos(k_{a}d)$ and $\Gamma^{B}=\Gamma_{_{1D}}+\Gamma'+\Gamma_{_{1D}}\cos(k_{a}d)$ are the decay rates of two dressed states $|A\rangle=(-|e\text{g}\rangle+|\text{g}e\rangle)/\sqrt{2}$ and $|B\rangle=(|e\text{g}\rangle+|\text{g}e\rangle)/\sqrt{2}$, respectively. Equations (\ref{eqa3a}) and (\ref{eqa3b}) show that, for a fixed atomic separation $d$, the decay rates $\Gamma^{A}$ and $\Gamma^{B}$ of the two dressed states are constant, while their energy levels can be tuned by controlling the characteristic strength $J$ and length $L$.

\section{Field response in the weak-excitation limit} \label{body}

\subsection{Energy shifts and effective decay rates} \label{energyshift}

In the single-excitation subspace, we diagonalize the effective non-Hermitian Hamiltonian $H_{1}={{\sum_{j=1}^{2}}}E_{j}|\psi_{j}^{R}\rangle\langle\psi_{j}^{L}|$ with $\langle\psi_{j}^{L}|\psi_{j'}^{R}\rangle=\delta_{jj'}$,
and obtain
\begin{eqnarray}            
E_{1}=-\Delta+J+Je^{-d/L}-\frac{\Gamma_{_{1D}}}{2}\sin(k_{a}d)-i\frac{\Gamma^{A}}{2},\label{eqa4a}\\
E_{2}=-\Delta+J-Je^{-d/L}+\frac{\Gamma_{_{1D}}}{2}\sin(k_{a}d)-i\frac{\Gamma^{B}}{2},\label{eqa4b}\\
|\psi_{1}^{R}\rangle=|A\rangle,\;\;\;\;\;\;\;\;|\psi_{2}^{R}\rangle=|B\rangle.\label{eqa4c}\;\;\;\;\;\;\;\;\;\;\;\;\;\;\;\;\;\;\;\;
\end{eqnarray}
With the eigenvalues and eigenmodes, we can write the transmission and reflection amplitudes of the incident photon as \cite{Shitao2015,Nie2021praapp}
\begin{eqnarray}           
t&=&1+\frac{i\Gamma_{_{1D}}}{2}{{\sum\limits_{j}}}\frac{V^{\dag}|\psi_{j}^{R}\rangle\langle\psi_{j}^{L}|V}{E_{j}},\label{eqa5a} \\
r&=&\frac{i\Gamma_{_{1D}}}{2}{{\sum\limits_{j}}}\frac{V^{\top}|\psi_{j}^{R}\rangle\langle\psi_{j}^{L}|V}{E_{j}},\label{eqa5b}
\end{eqnarray}
where the vector $V=(e^{ik_{a}x_{1}},e^{ik_{a}x_{2}})^{\top}$ denotes traveling photons in the 1D waveguide. In fact,
$t$ and $r$ can also be obtained with the input-output method \cite{CanevaNJP2015,Song2018}.
Equations (\ref{eqa5a}) and (\ref{eqa5b}) reveal that both the transmission and reflection amplitudes result from quantum interference between different scattering channels, which are determined by the energies and decay rates of the eigenstates.

\begin{figure}         
\centering\includegraphics[width=6.95 cm,angle=0]{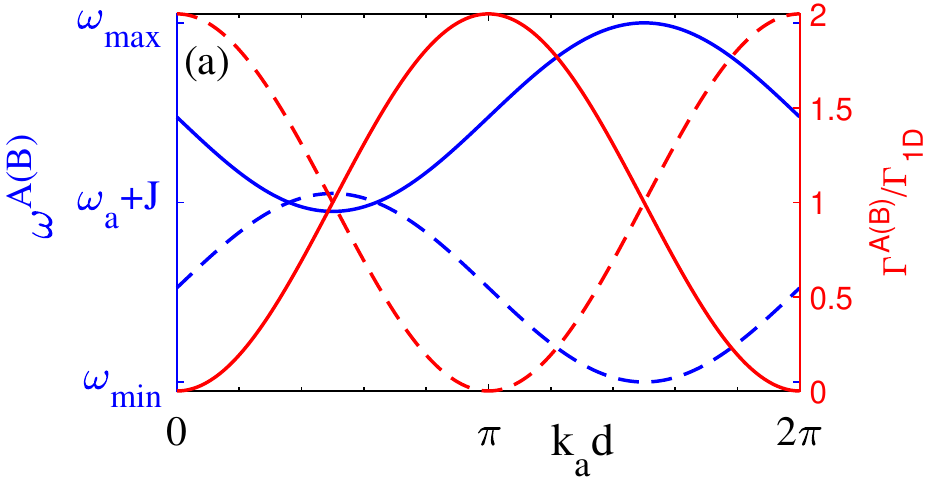}
\centering\includegraphics[width=7.0 cm,angle=0]{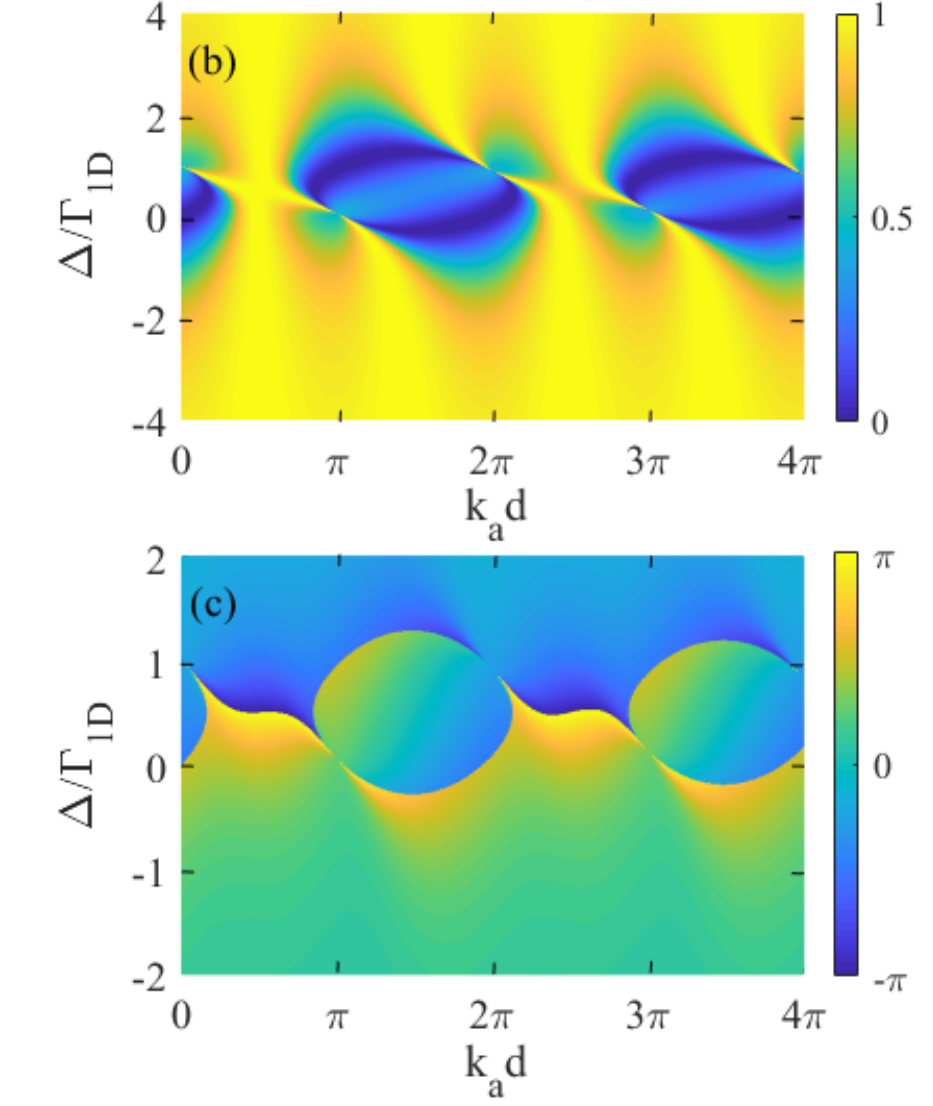}
\caption{ (a) The transition frequencies $\omega^{A}$ (blue solid line) and $\omega^{B}$ (blue dashed line) and the decay rates $\Gamma^{A}$ (red solid line) and $\Gamma^{B}$ (red dashed line) as a function of the parameter $k_{a}d$, with the labels $\omega_{\text{min}}=\omega_{a}+J-Je^{-d/L}-\Gamma_{_{1D}}/2$ and $\omega_{\text{max}}=\omega_{a}+J+Je^{-d/L}+\Gamma_{_{1D}}/2$. (b) The transmission $T$ and (c) phase shift $\theta$ of the input field as a function of the parameter $k_{a}d$ and detuning $\Delta/\Gamma_{_{1D}}$ with $J=0.5\Gamma_{_{1D}}$. In (a)-(c) $L=10\pi/k_{a}$, and $\Gamma'=0$.}
\label{figure3}
\end{figure}

In the weak-excitation regime, we give the energy levels and their decay rates into the waveguide modes in the
basis $\{$$|A\rangle$, $|B\rangle$$\}$ for three cases, i.e., general $d$, the Bragg case $k_{a}d=\pi$ and the anti-Bragg case $k_{a}d=0.5\pi$, as shown in Fig. \ref{figure2}. For general $d$, Eqs. (\ref{eqa3a}) and (\ref{eqa3b}) show that the dressed state $|A\rangle$ ($|B\rangle$) can be directly driven from the ground state $|G\rangle$ by a weakly exciting guided mode with transition frequency $\omega^{A}$ ($\omega^{B}$) and decay rate $\Gamma^{A}$ ($\Gamma^{B}$). Moreover, the energy levels of states $|A\rangle$ and $|B\rangle$ are shifted by $\Delta_{_{AB}}$. In fact, the distance $d$ between the two atoms affects the transition frequencies $\omega^{A}$ and $\omega^{B}$, as well as the decay rates $\Gamma^{A}$ and $\Gamma^{B}$, as shown in Fig. \ref{figure3}(a). For the regions $k_{a}d\in[0,\frac{\pi}{2})\cup(\frac{3\pi}{2},2\pi]$, state $|A\rangle$ ($|B\rangle$) is a subradiant (superradiant) state. However, for $k_{a}d\in(\frac{\pi}{2},\frac{3\pi}{2})$, state $|A\rangle$ ($|B\rangle$) becomes a superradiant (subradiant) state. Especially, in the Bragg case $k_{a}d=\pi$ with $\Gamma'=0$, we can get $\Gamma^{A}=2\Gamma_{_{1D}}$ and $\Gamma^{B}=0$. Intuitively, this phenomenon is due to the quantum interference between the emissions from the two atoms. That is, with $\Gamma'=0$, the relative phase in the $|A\rangle$ ($|B\rangle$) state lets the photon fields from the two atoms interfere constructively (destructively), which results in its decay rate being twice as strong (completely suppressed).

As shown in Fig. \ref{figure3}(b), we give the transmission $T$ of the input field traveling through
the two atoms with $J=0.5\Gamma_{_{1D}}$. The result reveals that, perfect transmission $T\approx1$
occurs for a large window even under the resonance condition $\Delta=0$. For example, in the anti-Bragg
case $k_{a}d=0.5\pi$ and $\Delta=0$, we can get $T=1$ with the conditions $2Je^{-d/L}=\Gamma_{_{1D}}$
and $\Gamma'=0$. This is very different from the single-qubit case \cite{ShenOL2005,Chang2007nap}
and the two-qubit case with no bound states \cite{zheng2013PRL}, where perfect transmission occurs
only for an off-resonance photon field. Intuitively, this phenomenon is caused by the long-range coherent
interactions between the two atoms arising from the bound states.
In fact,
under the condition $2Je^{-d/L}=\Gamma_{_{1D}}$ with $\Gamma'=0$, the transmission amplitude of the incident
photon for the anti-Bragg case can be written as
\begin{eqnarray}     \label{eqa5plus}       
t=\frac{\Delta-J-i\Gamma_{_{1D}}/2}{\Delta-J+i\Gamma_{_{1D}}/2}.
\end{eqnarray}
That is, in this case, perfect transmission $|t|=1$ occurs for an incident field with an arbitrary detuning $\Delta$,
which is due to the destructive quantum interference between the two scattering channels in Fig.
\ref{figure2}(c), yielding $r=0$. Moreover, we plot the phase shift
$\theta$ with the definition $t=\sqrt{T}e^{i\theta}$ in Fig. \ref{figure3}(c). The results
show that the phase shift is considerable and can be used to realize an atom-photon phase gate. Especially,
for the anti-Bragg case $k_{a}d=0.5\pi$, with the conditions $2Je^{-d/L}=\Gamma_{_{1D}}$, $\Delta=J$
and $\Gamma'=0$, we can obtain $T=1$ and $\theta=\pi$. In other words, the single photon travels through the
atom-waveguide system with a 100\% probability and a $\pi$ phase shift, which can be used to implement quantum phase gates \cite{Li2012prl}.

\subsection{The Bragg and anti-Bragg cases} \label{Braggcase}

\begin{figure}          
\centering
\includegraphics[width=6.5 cm,angle=0]{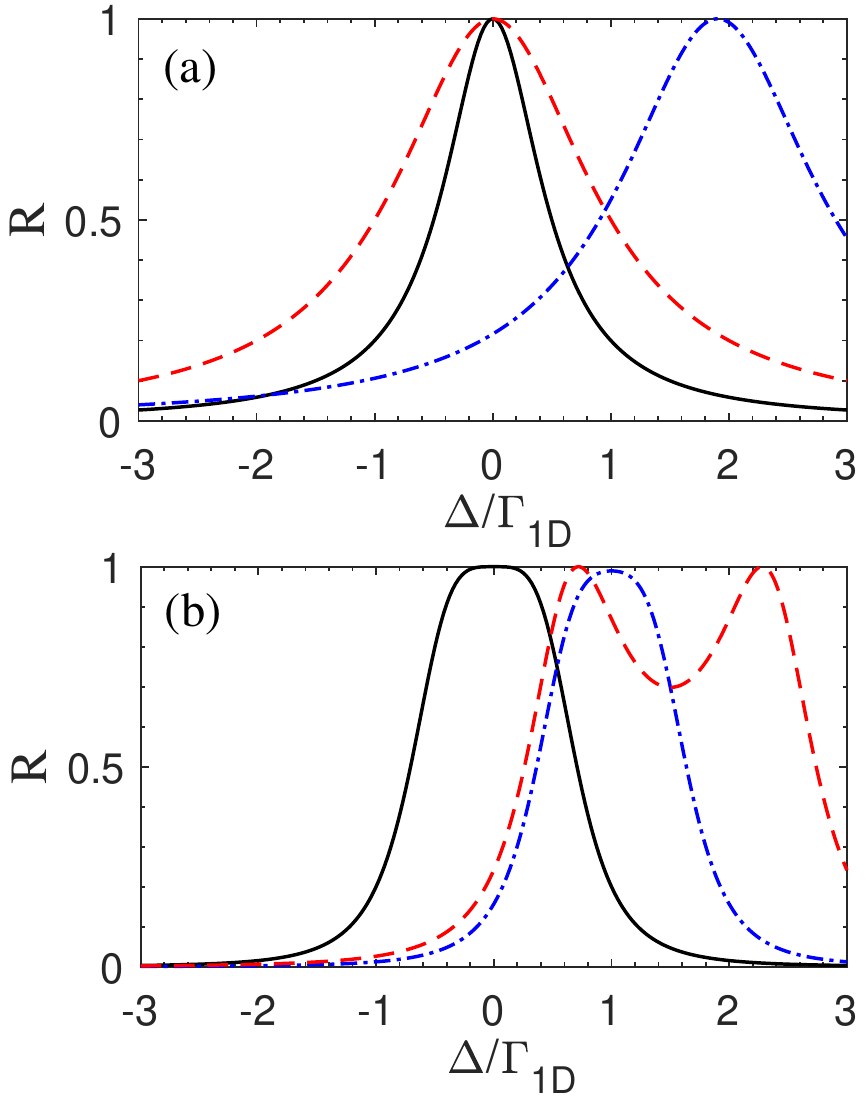}
\caption{ (a) The reflection $R$ of the incident field for a single atom (black solid line) and two atoms (red dashed line)
with no bound state and two atoms with $J=\Gamma_{_{1D}}$ (blue dash-dotted line) in the Bragg case $k_{a}d=\pi$.
(b) The reflection $R$ of the incident field for two atoms with $J=0$ (black solid line), $J=\Gamma_{_{1D}}$
(blue dash-dotted line), and $J=1.5\Gamma_{_{1D}}$ (red dashed line) in the anti-Bragg case $k_{a}d=0.5\pi$.
In (a) and (b) $L=10\pi/k_{a}$, and $\Gamma'=0$.  }
\label{figure4}
\end{figure}

As we discussed above, some interesting phenomena appear in the atom-waveguide system in the Bragg
and anti-Bragg cases, and we mainly focus on those two cases in the following sections. First, in the
Bragg case $k_{a}d=\pi$, Eqs. (\ref{eqa3a}) and (\ref{eqa3b}) become
\begin{eqnarray}
\nonumber          
H^{A}_{1}\!\!=\!\!\!\!\!\!&&(-\Delta\!+\!J\!+\!Je^{-d/L}\!-\!i\frac{\Gamma^{A}_{1}}{2})\sigma_{eg}^{A}\sigma_{ge}^{A}\!+\!\sqrt{2}\Omega_{p}(\sigma_{eg}^{A}\!+\!\sigma_{ge}^{A}),\label{eqa6a}\\
&&\\
H^{B}_{1}\!\!=\!\!\!\!\!\!&&(-\Delta\!+\!J\!-\!Je^{-d/L}\!-\!i\frac{\Gamma^{B}_{1}}{2})\sigma_{eg}^{B}\sigma_{ge}^{B},\label{eqa6b}
\end{eqnarray}
with $\Gamma^{A}_{1}=2\Gamma_{_{1D}}+\Gamma'$ and $\Gamma^{B}_{1}=\Gamma'$. For this case, the corresponding
energy levels and their decay rates into the waveguide modes in the basis $\{$$|A\rangle$, $|B\rangle$$\}$
are shown in Fig. \ref{figure2}(b). Equations (\ref{eqa6a}) and (\ref{eqa6b}) show that the superradiant state $|A\rangle$
(subradiant state $|B\rangle$) can (cannot) be driven from the ground state $|G\rangle$ by the weak driving field.
This results in a single peak in the reflection spectrum whose linewidth $\Gamma^{A}_{1}$ is larger than the
linewidth $\Gamma_{_{1D}}+\Gamma'$ in the single-atom case, as shown in Fig. \ref{figure4}(a). Moreover, the emission
of state $|B\rangle$ is perfectly suppressed with no atom dephasing and $\Gamma'=0$, which results in a
state in the decoherence-free subspace. Compared to the two-atom case with no bound state \cite{zheng2013PRL},
the reflection spectrum in our model is shifted by $J+Je^{-d/L}$, as shown in Fig. \ref{figure4}(a).

However, for the anti-Bragg case $k_{a}d=0.5\pi$, Eqs. (\ref{eqa3a}) and (\ref{eqa3b}) evolve into
\begin{eqnarray}            
\nonumber
H^{A}_{2}=\!\!\!\!&&\big[-\Delta+J+(Je^{-d/L}-\frac{\Gamma_{_{1D}}}{2})-i\frac{\Gamma^{A}_{2}}{2}\big]\sigma_{eg}^{A}\sigma_{ge}^{A}\\
&&+\frac{\Omega_{p}}{\sqrt{2}}\big[(1-i)\sigma_{eg}^{A}+\mathrm{H.c.}\big],\label{eqa7a}\\\nonumber
H^{B}_{2}=\!\!\!\!&&\big[-\Delta+J-(Je^{-d/L}-\frac{\Gamma_{_{1D}}}{2})-i\frac{\Gamma^{B}_{2}}{2}\big]\sigma_{eg}^{B}\sigma_{ge}^{B}\\
&&-\frac{\Omega_{p}}{\sqrt{2}}\big[(1+i)\sigma_{eg}^{B}+\mathrm{H.c.}\big],\label{eqa7b}
\end{eqnarray}
with $\Gamma^{A}_{2}=\Gamma_{_{1D}}+\Gamma'$ and $\Gamma^{B}_{2}=\Gamma_{_{1D}}+\Gamma'$. Figure \ref{figure2}(c)
shows the corresponding energy levels and their decay rates into the waveguide modes in dressed states
$|A\rangle$ and $|B\rangle$. Equations (\ref{eqa7a}) and (\ref{eqa7b}) reveal that both $|A\rangle$ and $|B\rangle$ can be driven
from the ground state $|G\rangle$ by the weak input field with equal amplitudes. The energy difference between
the two dressed states is $2Je^{-d/L}-\Gamma_{_{1D}}$, and the linewidths of both states are $\Gamma_{_{1D}}+\Gamma'$,
meaning that we can distinguish the two resonances in the reflection spectrum only when $2Je^{-d/L}-\Gamma_{_{1D}}>\Gamma_{_{1D}}+\Gamma'$.
In Fig. \ref{figure4}(b), we give the reflection spectrum of the input field in three cases, i.e., $J=0$, $J=\Gamma_{_{1D}}$, and $J=1.5\Gamma_{_{1D}}$.
The results show that, in the cases with $J=0$ and $J=\Gamma_{_{1D}}$, satisfying $2Je^{-d/L}-\Gamma_{_{1D}}<\Gamma_{_{1D}}+\Gamma'$, there is a single peak in the reflection spectrum.
However, a two-peak reflection spectrum can be obtained for the case $J=1.5\Gamma_{_{1D}}$, which satisfies the condition $2Je^{-d/L}-\Gamma_{_{1D}}>\Gamma_{_{1D}}+\Gamma'$. Thus, for the anti-Bragg case, in the two-atom case with no bound states i.e., $J=0$ \cite{zheng2013PRL}, we can observe only a single-peak reflection spectrum even though the system has two resonance modes. Moreover, in the anti-Bragg case, when $2Je^{-d/L}=\Gamma_{_{1D}}$, dressed states $|A\rangle$ and $|B\rangle$ with equal decay rates have the same energy, and destructive quantum interference occurs in the two scattering channels shown in Fig. \ref{figure2}(c) for the photon reflection process, yielding $r=0$ even when $\Gamma'\neq0$, which can be derived from Eq. (\ref{eqa5b}).

\begin{figure}          
\centering
\includegraphics[width=6.5 cm,angle=0]{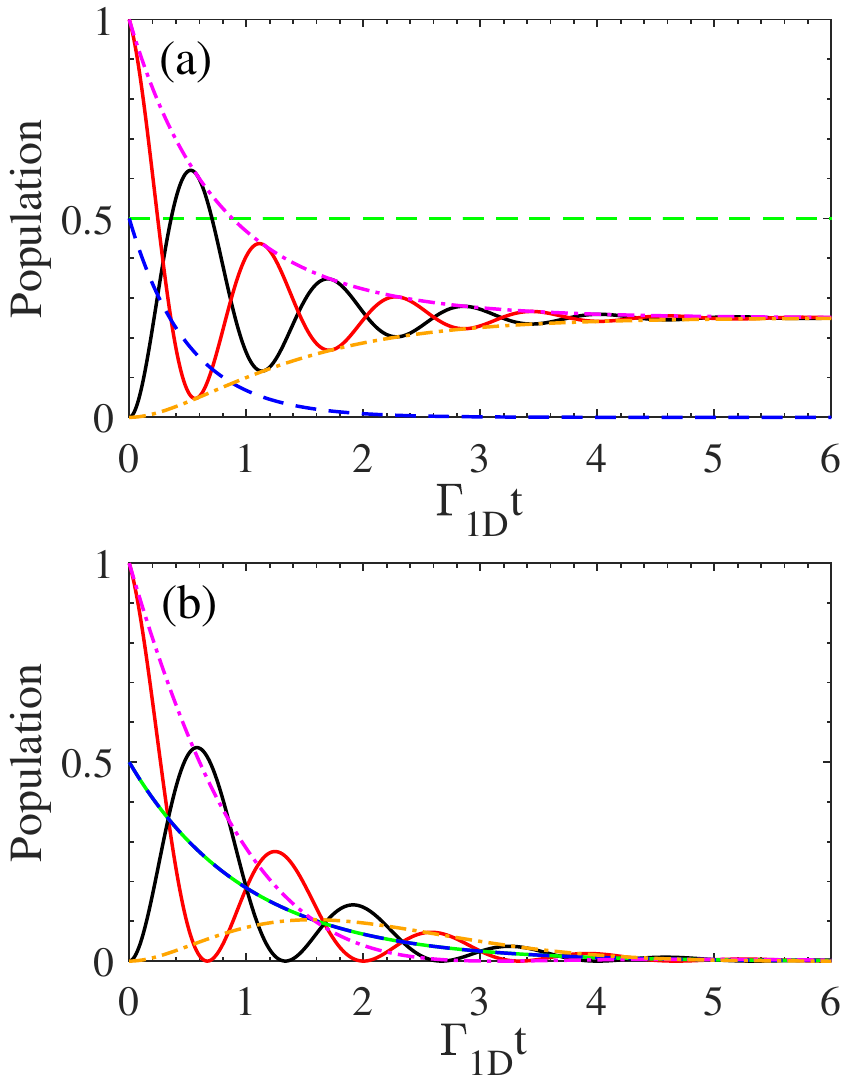}
\caption{Evolution of the excited-state populations of the left atom (magenta dash-dotted line for $J=0$, red solid line for $J=3\Gamma_{_{1D}}$) and right atom (orange dash-dotted line for $J=0$, black solid line for $J=3\Gamma_{_{1D}}$) and the populations of the two dressed states $|A\rangle$ (blue dashed line) and $|B\rangle$ (green dashed line) for the (a) Bragg case $k_{a}d=\pi$ and (b) anti-Bragg case $k_{a}d=0.5\pi$ when the left atom is excited at the initial time. In (a) and (b) $L=10\pi/k_{a}$, $\Omega_{p}=0$, and $\Gamma'=0$.}
\label{figure5}
\end{figure}

In Fig. \ref{figure5}, with no driving field ($\omega_{p}=0$, $\Omega_{p}=0$),  we give the evolution of the excited-state populations of the two atoms and the populations of dressed states $|A\rangle$ and $|B\rangle$ for the Bragg and anti-Bragg cases when the left atom is initially excited. In fact, for the Bragg case, the excited-state populations of the left and right atoms and the populations of states $|A\rangle$ and $|B\rangle$ can be written as
\begin{eqnarray}       
p^{L}_{1}\!\!\!&=&\!\!\!\frac{1}{4}e^{-(\Gamma_{_{1D}}+\Gamma')t}[e^{\Gamma_{_{1D}}t}\!+\!e^{-\Gamma_{_{1D}}t}\!+\!2\cos(2Je^{-d/L}t)],\;\;\;\;\;\;\label{eqa8a}\\
p^{R}_{1}\!\!\!&=&\!\!\!\frac{1}{4}e^{-(\Gamma_{_{1D}}+\Gamma')t}[e^{\Gamma_{_{1D}}t}\!+\!e^{-\Gamma_{_{1D}}t}\!-\!2\cos(2Je^{-d/L}t)],\;\;\;\;\;\;\label{eqa8b}\\
p^{A}_{1}\!\!\!&=&\!\!\!\frac{1}{2}e^{-(2\Gamma_{_{1D}}+\Gamma')t},\;\;\;\;\;\;\;\;p^{B}_{1}=\frac{1}{2}e^{-\Gamma't}.\label{eqa8c}
\end{eqnarray}
Figure \ref{figure5}(a) and Eqs. (\ref{eqa8a}) and (\ref{eqa8b}) show that, in the Bragg case, the existence of the bound states causes the oscillations in the populations of the left and right atoms, which vanish for $J=0$. However, the bound states have no influence on the populations of states $|A\rangle$ and $|B\rangle$.
Interestingly, with $\Gamma'=0$, the excited-state populations of the left and right atoms are both $25\%$ when time is infinite and the population of the $|B\rangle$ state is $50\%$ at any time, which does not depend on the existence of bound states.

\begin{figure}          
\centering
\includegraphics[width=6.6 cm,angle=0]{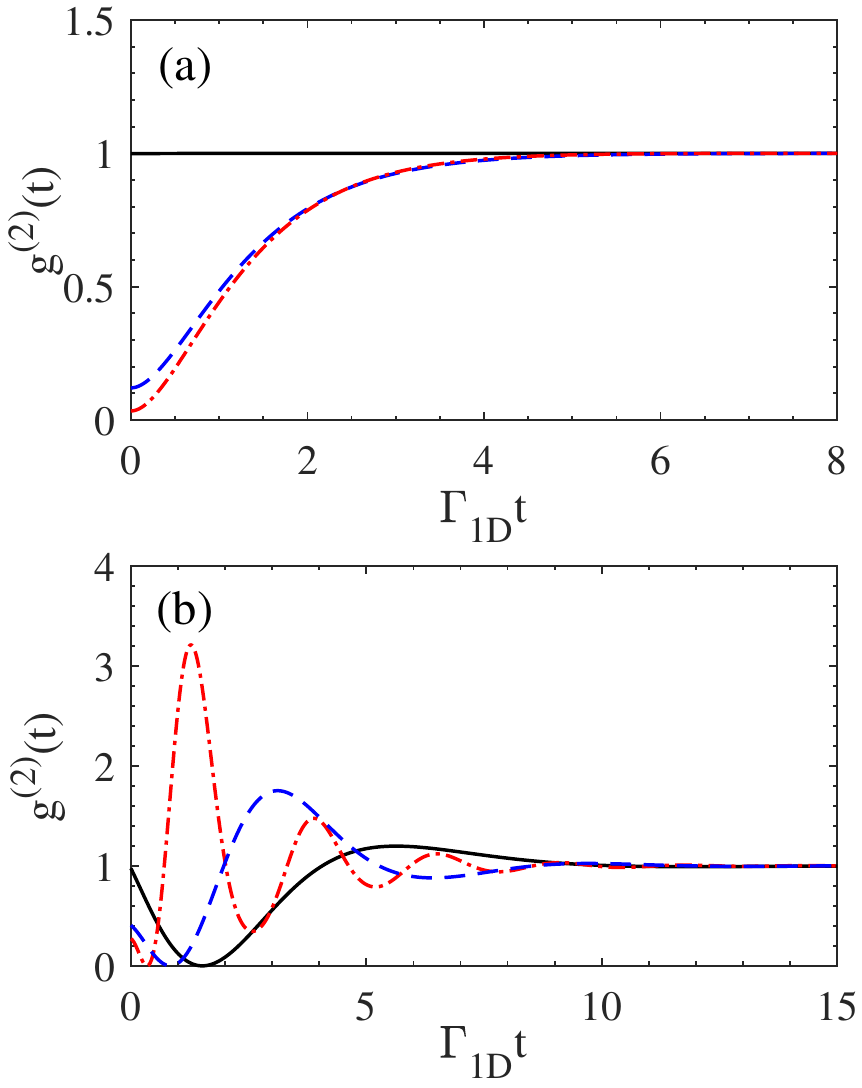}
\caption{The photon-photon correlation function $\text{g}^{(2)}(t)$ of the reflected field with $J=0$ (black solid line), $J=1.5\Gamma_{_{1D}}$ (blue dashed line), and $J=3\Gamma_{_{1D}}$ (red dashed-dotted line) for the (a) Bragg case $k_{a}d=\pi$ with $\Delta=J+Je^{-d/L}$ and (b) anti-Bragg case $k_{a}d=0.5\pi$ with $\Delta=J$. In
(a) and (b) $L=10\pi/k_{a}$, $\Omega_{p}=0.0001\Gamma_{_{1D}}$, and $\Gamma'=0$.}
\label{figure6}
\end{figure}

For the anti-Bragg case, the four populations become
\begin{eqnarray}           
p^{L}_{2}\!\!&=&\!\!\frac{1}{4}e^{-(\Gamma_{_{1D}}+\Gamma')t}\big\{2+2\cos[(2Je^{-d/L}-\Gamma_{_{1D}})t]\big\},\;\;\;\;\;\;\label{eqa9a} \\
p^{R}_{2}\!\!&=&\!\!\frac{1}{4}e^{-(\Gamma_{_{1D}}+\Gamma')t}\big\{2-2\cos[(2Je^{-d/L}-\Gamma_{_{1D}})t]\big\},\;\;\;\;\;\;\label{eqa9b} \\
p^{A}_{2}\!\!&=&\!\!p^{B}_{2}=\frac{1}{2}e^{-(\Gamma_{_{1D}}+\Gamma')t}.
\end{eqnarray}
As shown in Fig. \ref{figure5}(b)and Eqs. (\ref{eqa9a}) and (\ref{eqa9b}), even with $J=0$, the oscillations also appear in the excited-state populations of the left and right atoms. However, the oscillation phenomenon disappears under the condition $2Je^{-d/L}=\Gamma_{_{1D}}$, which can easily be derived from Eqs. (\ref{eqa9a}) and (\ref{eqa9b}). In this case, the left atom decays with a rate $\Gamma_{_{1D}}+\Gamma'$, and the right atom is not excited at any time, which indicates that the left and right atoms no longer interact with each other. This can be interpreted by the non-Hermitian Hamiltonian in Eq. (\ref{eqa1}), which becomes $H'_{1}=[\omega_{a}+J-\frac{i(\Gamma_{_{1D}}+\Gamma')}{2}](\sigma_{ee}^{1}+\sigma_{ee}^{2})$ under the condition $2Je^{-d/L}=\Gamma_{_{1D}}$ for the anti-Bragg case with no probe field. In addition, the populations of dressed states $|A\rangle$ and $|B\rangle$ are the same at any time, decay to zero with a rate $\Gamma_{_{1D}}+\Gamma'$, and are independent of the bound states.

\subsection{Photon-photon correlation}

As we know, the main feature of nonclassical light can be revealed by the photon-photon correlation function (second-order correlation function) \cite{Loudon2003,Sanche2020prl}.
Assuming a weak continuous coherent input field in our system, we give the photon-photon correlation function $\text{g}^{(2)}(t)$ of the reflected field for the Bragg and anti-Bragg cases in Fig. \ref{figure6}. The results show that, for the two cases, initial antibunching ($\text{g}^{(2)}<1$) appears in the photon-photon correlation of the reflected field with $J\neq0$. For the Bragg case $k_{a}d=\pi$, as shown in Fig. \ref{figure6}(a), only antibunching can be observed,  and no oscillation occurs in the photon-photon correlation of the reflected field, which is similar to that for a single qubit with no bound state. This can be interpreted by the energy levels in the basis $\{$$|A\rangle$, $|B\rangle$$\}$ for the Bragg case shown in Fig. \ref{figure2}(b). In this case, only the superradiant state $|A\rangle$ is coupled to the waveguide modes, which is equivalent to a single two-level qubit. However, for the anti-Bragg case, the two dressed states, $|A\rangle$ and $|B\rangle$, are both coupled to the waveguide modes with equal decay rates $\Gamma_{_{1D}}+\Gamma'$. Thus, the quantum interference between the transitions $|A\rangle\rightarrow|G\rangle$ and $|B\rangle\rightarrow|G\rangle$ results in oscillations between antibunching ($\text{g}^{(2)}<1$) and bunching ($\text{g}^{(2)}>1$), as shown in Fig. \ref{figure6}(b). Moreover, the frequency of the oscillation is $\omega_{_{A}}-\omega_{_{B}}=2Je^{-d/L}-\Gamma_{_{1D}}$, which can be tuned by the control of the parameters $J$ and $L$. Specifically, under the condition $2Je^{-d/L}=\Gamma_{_{1D}}$, quantum beats disappear in the photon-photon correlation function.

\subsection{Directional photon emission}

To proceed, we study directional photon emission in our system in the anti-Bragg case $k_{a}d=0.5\pi$.
With the input-output relations \cite{klalum2013}, the output rightward- and leftward-propagating modes in the waveguide can be written as
\begin{eqnarray}     \label{eqa9aplus}
a_{r}(x)&=& a_{r}^{in} +\frac{i\Gamma_{_{1D}}}{2}e^{ik_{a}x}(\sigma_{ge}^{1}+\sigma_{ge}^{2}e^{-ik_{a}d}),\\
a_{l}(x)&=& a_{l}^{in} +\frac{i\Gamma_{_{1D}}}{2}e^{-ik_{a}x}(\sigma_{ge}^{1}+\sigma_{ge}^{2}e^{ik_{a}d}).
\end{eqnarray}
Here, $a_{r}^{in}$ and $a_{l}^{in}$ denote the rightward- and leftward-propagating weak input fields in the waveguide, respectively. Thus, the number of photons in either mode of the waveguide can be defined as $\langle \hat{n}_{r(l)} \rangle = \langle \psi(t)|a_{r(l)}^{\dag} a_{r(l)}| \psi(t) \rangle$, where $|\psi(t)\rangle=\exp(-iHt)|\psi(0)\rangle$ is the many-body wave function of the
system, with $|\psi(0)\rangle$ being the initial state. Here, we define two new dressed-state operators,
\begin{eqnarray}     \label{eqa9bplus}       
\sigma_{e\text{g}}^{A'}=\frac{\sigma_{e\text{g}}^{1}-i\sigma_{e\text{g}}^{2}}{\sqrt{2}},\;\;\;\;\;\sigma_{e\text{g}}^{B'}=\frac{\sigma_{e\text{g}}^{1}+i\sigma_{e\text{g}}^{2}}{\sqrt{2}},
\end{eqnarray}
where the two new dressed states are $|A'\rangle=(|e\text{g}\rangle-i|\text{g}e\rangle)/\sqrt{2}$ and $|B'\rangle=(|e\text{g}\rangle+i|\text{g}e\rangle)/\sqrt{2}$, respectively.

First, we focus on the case in which the two atoms are initialized to $|A'\rangle$ or $|B'\rangle$ and no probe field is input into the waveguide, i.e., $a_{r}^{in}=a_{l}^{in}=0$. In detail, when the two atoms are initially prepared in $|A'\rangle$ ($|B'\rangle$), we get $\langle \hat{n}_{r} \rangle=0$ ($\langle \hat{n}_{l} \rangle=0$), which indicates that the dimer emits only a leftward-propagating (rightward-propagating) single photon.
Taking the case $|A'\rangle$ as an example, such a phenomenon results from the full constructive (destructive) interference between
the two pathways for the photon field emitted by the two atoms into the left-(right-) propagating modes of the 1D PCW.

\begin{figure}          
\centering
\includegraphics[width=6.8 cm,angle=0]{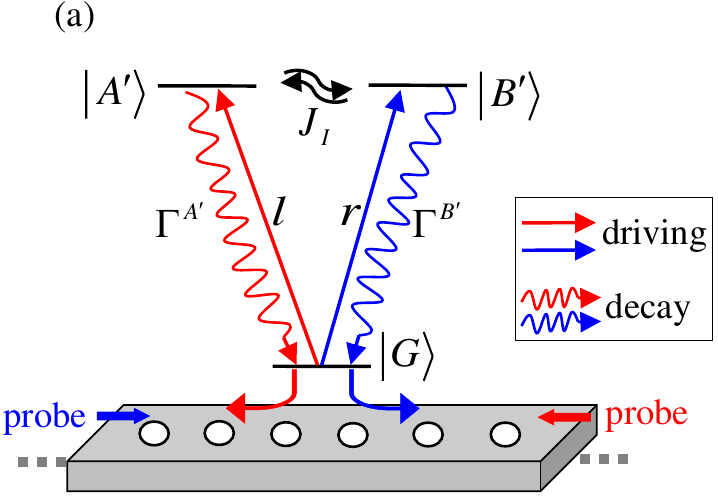}
\includegraphics[width=7.5 cm,angle=0]{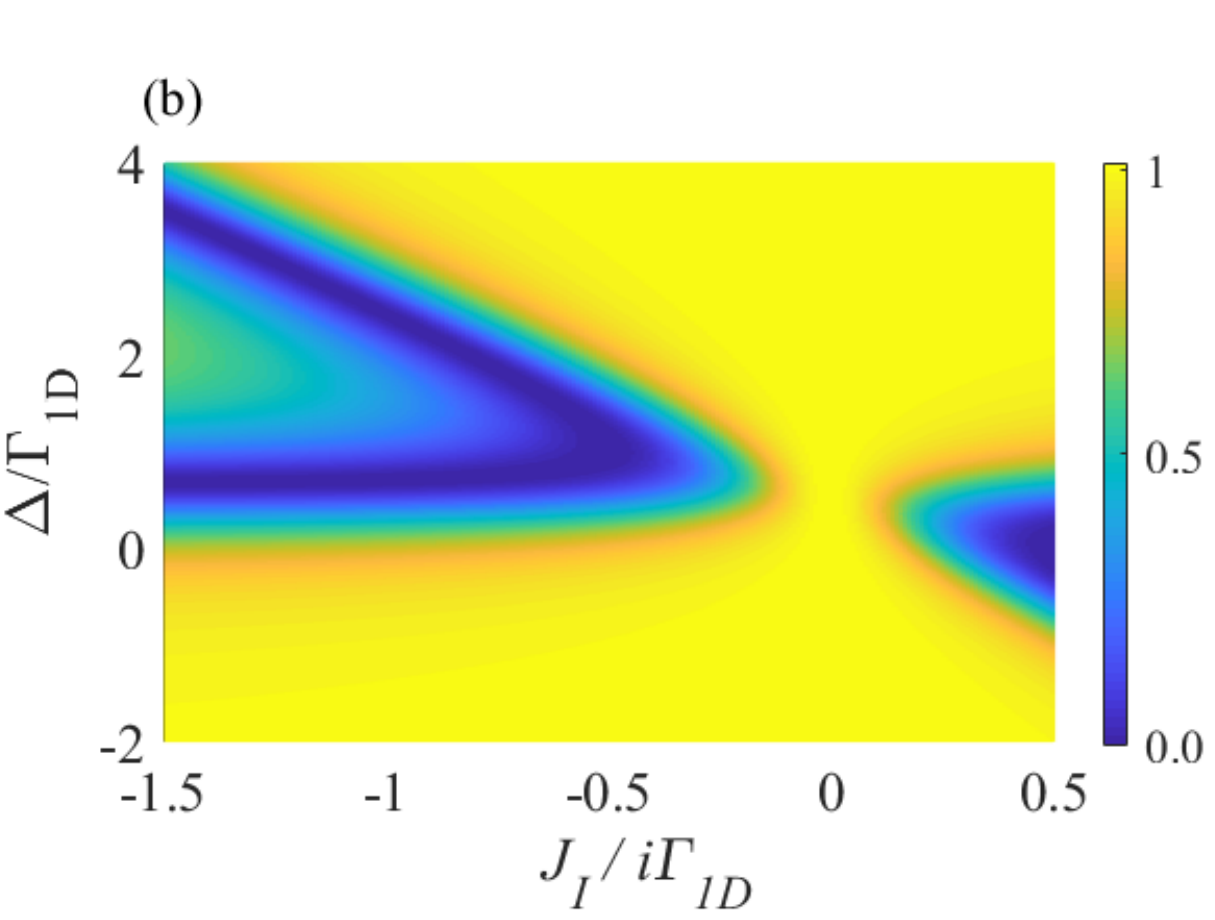}
\caption{(a) Energy levels and their decay rates into the waveguide modes in the new dressed basis $\{$$|A'\rangle$, $|B'\rangle$$\}$ for the anti-Bragg case $k_{a}d=0.5\pi$. $|ee\rangle$ is ignored for weak probe fields. The transition $|G\rangle\leftrightarrow|A'\rangle$ can be driven only by a leftward-propagating probe field, and state $|A'\rangle$ can emit only a leftward-propagating photon (red).  The transition $|G\rangle\leftrightarrow|B'\rangle$ can be driven only by a rightward-propagating probe field, and state $|B'\rangle$ can emit only a rightward-propagating photon (blue). $\Gamma^{A'}=\Gamma^{B'}=\Gamma_{_{1D}}+\Gamma'$, and $J_{I}=i(\Gamma_{_{1D}}/{2}-Je^{-d/L})$.
(b) The transmission $T$ of the probe field input from the left or right as a function of the parameter $J_{I}/i\Gamma_{_{1D}}$ and detuning $\Delta/\Gamma_{_{1D}}$, with $L=10\pi/k_{a}$ and $\Gamma'=0$ in the anti-Bragg case $k_{a}d=0.5\pi$. }
\label{Fig6plus}
\end{figure}

Next, we consider the case in which a weak probe field is input from the left or right.
For the anti-Bragg case $k_{a}d=0.5\pi$, if the weak probe field is input from the left, with $a_{l}^{in}=0$ and $a_{r}^{in}=\Omega_{p}e^{ik_{p}x}$,
the total Hamiltonian can be written as $H=H_{1}^{A'}+H_{1}^{B'}+H_{1}^{A'B'}$, with
\begin{eqnarray}
H_{1}^{A'}\!\!=\!\!\!\!\!\!&&(-\Delta\!+\!J\!-\!i\frac{\Gamma_{_{1D}}+\Gamma'}{2})\sigma_{eg}^{A'}\sigma_{ge}^{A'},\label{eqa9c1plus}\\
\nonumber
H_{1}^{B'}\!\!=\!\!\!\!\!\!&&(-\Delta\!+\!J\!-\!i\frac{\Gamma_{_{1D}}+\Gamma'}{2})\sigma_{eg}^{B'}\sigma_{ge}^{B'}\!-\!\sqrt{2}\Omega_{p}(\sigma_{eg}^{B'}\!+\!\mathrm{H.c.}),\\\label{eqa9c2plus}
&&\\
H_{1}^{A'B'}\!\!=\!\!\!\!\!\!&&i(\frac{\Gamma_{_{1D}}}{2}\!-\!Je^{-d/L})(\sigma_{eg}^{A'}\sigma_{ge}^{B'}-\mathrm{H.c.}).\label{eqa9c3plus}
\end{eqnarray}
Equations (\ref{eqa9c1plus}) and (\ref{eqa9c2plus}) show that the probe field input from the left drives only the atomic transition $|G\rangle\leftrightarrow|B'\rangle$.
However, if the weak probe field is incident from the right, with $a_{l}^{in}=\Omega_{p}e^{-ik_{p}x}$ and $a_{r}^{in}=0$, the corresponding driving Hamiltonian is $H_{2'}=-{{\sum\limits_{j,k=1}^2}}\Omega_{p}(e^{-ik_{p}x_{_{j}}}\sigma_{e\text{g}}^{j}+\mathrm{H.c.})$.
The total Hamiltonian can be rewritten as $H=H_{2}^{A'}+H_{2}^{B'}+H_{2}^{A'B'}$, with
\begin{eqnarray}
\nonumber
H_{2}^{A'}\!\!=\!\!\!\!\!\!&&(-\Delta\!+\!J\!-\!i\frac{\Gamma_{_{1D}}+\Gamma'}{2})\sigma_{eg}^{A'}\sigma_{ge}^{A'}\!-\!\sqrt{2}\Omega_{p}(\sigma_{eg}^{A'}+\mathrm{H.c.}),\label{eqa9d1plus}\\
&&\\
H_{2}^{B'}\!\!=\!\!\!\!\!\!&&(-\Delta\!+\!J\!-\!i\frac{\Gamma_{_{1D}}+\Gamma'}{2})\sigma_{eg}^{B'}\sigma_{ge}^{B'},\label{eqa9d2plus}\\
H_{2}^{A'B'}\!\!=\!\!\!\!\!\!&&i(\frac{\Gamma_{_{1D}}}{2}\!-\!Je^{-d/L})(\sigma_{eg}^{A'}\sigma_{ge}^{B'}\!-\!\mathrm{H.c.}).\label{eqa9d3plus}
\end{eqnarray}
Equations (\ref{eqa9d1plus}) and (\ref{eqa9d2plus}) reveal that the input field from the right drives only the atomic transition $|G\rangle\leftrightarrow|A'\rangle$.
Note that, in the above two cases, the dressed states $|A'\rangle$ and $|B'\rangle$ have the same energy and decay rate, and exchange interactions between the two states with strength $J_{I}=i(\Gamma_{_{1D}}/{2}-Je^{-d/L})$ exist. Under the condition $J_{I}=0$, the input field propagating toward the right (left) drives only the atomic transition $|G\rangle\leftrightarrow|B'\rangle$ ($|G\rangle\leftrightarrow|A'\rangle$), and state $|B'\rangle$ ($|A'\rangle$) can reemit only a rightward-propagating (leftward-propagating) photon field. That is, the direction of the photon field re-emitted by the two dressed states is the same as the input probe field, as shown in Fig. \ref{Fig6plus}(a). This interesting phenomenon results in perfect transmission for the weak input field. However, if $J_{I}\neq0$, population transfer between the two states will occur, which
results in part of the atomic emission traveling in the direction opposite that of the input probe field. Thus, in the case with $J_{I}\neq0$, perfect transmission is destroyed by the population transfer between states $|A'\rangle$ and $|B'\rangle$. To verify this, we calculate the transmission of the weak probe field input from the left or right as a function of the parameter $J_{I}/i\Gamma_{_{1D}}$ and detuning $\Delta/\Gamma_{_{1D}}$, as shown in Fig. \ref{Fig6plus}(b).
The results show that, for $|J_{I}|>\Gamma_{_{1D}}/2$, i.e., $Je^{-d/L}>\Gamma_{_{1D}}$, we observe two dips in
the transmission which are related to the hybridized energy splitting of the dressed states $|A'\rangle$ and $|B'\rangle$.
In fact, this phenomenon corresponds to the results shown in Fig. \ref{figure4}(b).
In addition, as mentioned above, we see that the transmission of the probe field approaches unity for any detuning $\Delta$ when $J_{I}=0$, i.e., the condition $2Je^{-d/L}=\Gamma_{_{1D}}$, which is consistent with the results
given in Eq. (\ref{eqa5plus}).

\subsection{Effects of the imperfections in realistic atom-waveguide systems} \label{imperfection}

\begin{figure}          
\centering
\includegraphics[width=6.5 cm,angle=0]{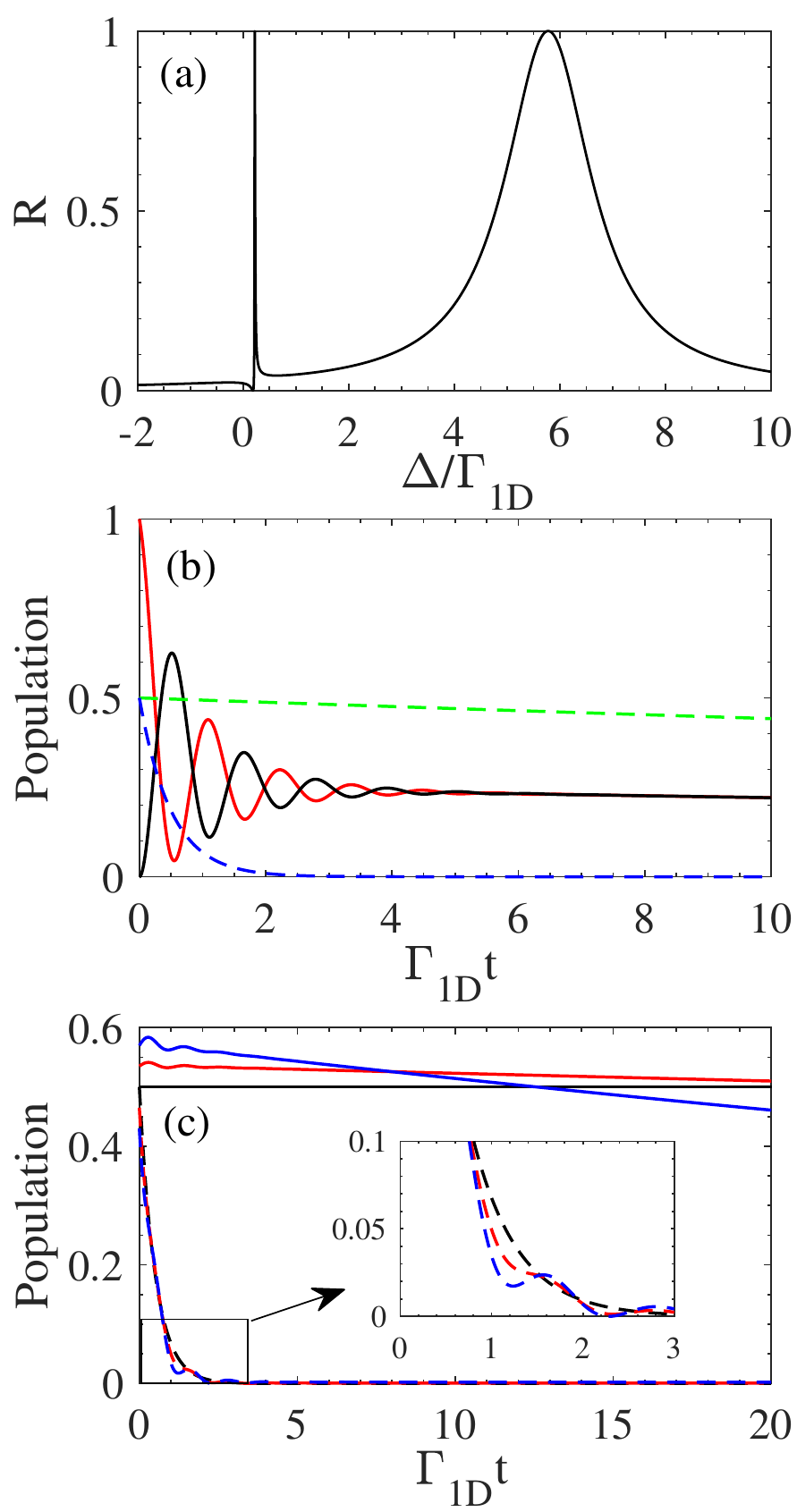}
\caption{(a) The reflection spectrum of our two-atom system. (b) Evolution of the excited-state populations of the left atom (red solid line) and right atom (black solid line) and the populations of the dressed states $|A\rangle$ (blue dashed line) and $|B\rangle$ (green dashed line) when the left atom is excited at the initial time with no driving field. (c) The populations of states $|A''\rangle$ (dashed lines) and $|B''\rangle$ (solid lines) for $\xi=0$ (black lines), $\xi=0.07$ (red lines), and $\xi=0.14$ (blue lines), when the left atom is excited at the initial time with no driving field. In (a) and (b) $\eta=0.05$ and in (a)-(c) $J=3\Gamma_{_{1D}}$, $L=10\pi/k_{a}$, and $\Gamma'=0$.}
\label{figure7}
\end{figure}

In the above discussion, we mainly focused on the Bragg case $k_{a}d=\pi$ and anti-Bragg case $k_{a}d=0.5\pi$. However, in realistic waveguide systems, the distance between the two atoms may deviate from these specific values. Now, we study the effects of deviation in the distance $d$ between two atoms in our system, taking the Bragg case as an example. Here, we set $k_{a}d=(1+\eta)\pi$, where $|\eta|\ll1$ is the deviation parameter. In this case, with $\eta\neq0$, a sharp asymmetric peak, called Fano resonance, can be observed in the reflection spectrum, as shown in Fig. \ref{figure7}(a).
This Fano resonance is caused by the subradiant state $|B\rangle$, which can be driven directly from state $|G\rangle$, and its decay rate $\Gamma^{B}$ is very small but not zero when $\eta\neq0$. The Fano resonance is useful for optical sensing and switching \cite{Limonov2017NP}. For example, the Fano resonance occurs around $\Delta=J-Je^{-d/L}+\Gamma_{_{1D}}\sin(k_{a}d)/2$ in our system, which can be used to estimate the characteristic strength $J$ and length $L$. The above results reveal that the reflection spectrum is very sensitive to the deviation parameter. Moreover, in Fig. \ref{figure7}(b), with no driving field ($\Omega_{p}=0$), we give the evolution of the excited-state populations of two atoms and the populations of the two dressed states $|A\rangle$ and $|B\rangle$ for the case $k_{a}d=(1+\eta)\pi$ when the left atom is initially excited.
The results show that, with $\eta\neq0$ and $\Gamma'=0$, the populations of the two atoms in their excited states and state $|B\rangle$ both decay to zero very slowly, rather than holding at specific values when the time is infinite in the Bragg case. This is also due to the small, but nonzero, decay rate of the subradiant state $|B\rangle$.
The three populations are not sensitive to the deviation parameter. In fact, the coupling points of superconducting artificial atoms in a waveguide can be manipulated with high precision in experiments \cite{Kannan2020na}, which means that the deviation of the distance can be strongly suppressed.

So far, we have assumed that the decay rates to the waveguide modes for the two atoms are the same, i.e., $\Gamma_{_{1D}}$. However, for a realistic waveguide system,
the decay rates for the two atoms may be different and are given by $\Gamma_{_{1D,1}}\neq\Gamma_{_{1D,2}}$ \cite{Mirhos2019}. In this case, we define two new dressed states, $|A''\rangle$ and $|B''\rangle$, as
\begin{eqnarray}           
|A''\rangle&=&\frac{-\sqrt{\Gamma_{_{1D,1}}}|e\text{g}\rangle+\sqrt{\Gamma_{_{1D,2}}}|\text{g}e\rangle}{\sqrt{\Gamma_{_{1D,1}}+\Gamma_{_{1D,2}}}},\label{eqa10a} \\
|B''\rangle&=&\frac{\sqrt{\Gamma_{_{1D,2}}}|e\text{g}\rangle+\sqrt{\Gamma_{_{1D,1}}}|\text{g}e\rangle}{\sqrt{\Gamma_{_{1D,1}}+\Gamma_{_{1D,2}}}}.\label{eqa10b}
\end{eqnarray}
The two dressed-state operators in Eq. (\ref{eqa2}) are changed into
\begin{eqnarray}            
\sigma_{e\text{g}}^{A''}&=&\frac{-\sqrt{\Gamma_{_{1D,1}}}\sigma_{e\text{g}}^{1}+\sqrt{\Gamma_{_{1D,2}}}\sigma_{e\text{g}}^{2}}{\sqrt{\Gamma_{_{1D,1}}+\Gamma_{_{1D,2}}}},\label{eqa11a}\\
\sigma_{e\text{g}}^{B''}&=&\frac{\sqrt{\Gamma_{_{1D,2}}}\sigma_{e\text{g}}^{1}+\sqrt{\Gamma_{_{1D,1}}}\sigma_{e\text{g}}^{2}}{\sqrt{\Gamma_{_{1D,1}}+\Gamma_{_{1D,2}}}}.\label{eqa11b}
\end{eqnarray}
Since we observe a dark state only in the Bragg case, we mainly focus here on the effect of asymmetry in the decay rate $\Gamma_{_{1D}}$ in this case.
With the new definitions, in the Bragg case, the total Hamiltonian can be rewritten as $H = H^{A''} + H^{B''} + H^{A''B''}$, where
\begin{eqnarray}            
H^{A''}=&&\!\!\!\!\big[\!-\!\Delta\!+\!J\!-\!\frac{Je^{-d/L}(\Gamma_{_{1D,1}}-\Gamma_{_{1D,2}})^{2}}{2\sqrt{\Gamma_{_{1D,1}}\Gamma_{_{1D,2}}}(\Gamma_{_{1D,1}}+\Gamma_{_{1D,2}})}\nonumber\\
&&+Je^{-d/L}\frac{\Gamma_{_{1D,1}}+\Gamma_{_{1D,2}}}{2\sqrt{\Gamma_{_{1D,1}}\Gamma_{_{1D,2}}}}\!-\!i\frac{\Gamma^{A''}}{2}\big]\sigma_{eg}^{A''}\sigma_{ge}^{A''}\nonumber\\
&&+\;\Omega_{p}(\frac{\sqrt{\Gamma_{_{1D,1}}}+\sqrt{\Gamma_{_{1D,2}}}}{\sqrt{\Gamma_{_{1D,1}}+\Gamma_{_{1D,2}}}}\sigma_{eg}^{A''}+\mathrm{H.c.}),\label{eqa12a}\\
H^{B''}=&&\!\!\!\!\big[\!-\!\Delta\!+\!J\!+\frac{Je^{-d/L}(\Gamma_{_{1D,1}}-\Gamma_{_{1D,2}})^{2}}{2\sqrt{\Gamma_{_{1D,1}}\Gamma_{_{1D,2}}}(\Gamma_{_{1D,1}}+\Gamma_{_{1D,2}})}\nonumber\\
&&-Je^{-d/L}\frac{\Gamma_{_{1D,1}}+\Gamma_{_{1D,2}}}{2\sqrt{\Gamma_{_{1D,1}}\Gamma_{_{1D,2}}}}\!-\!i\frac{\Gamma^{B''}}{2}\big]\sigma_{eg}^{B''}\sigma_{ge}^{B''}\nonumber\\
&&+\;\Omega_{p}(\frac{\sqrt{\Gamma_{_{1D,1}}}-\sqrt{\Gamma_{_{1D,2}}}}{\sqrt{\Gamma_{_{1D,1}}+\Gamma_{_{1D,2}}}}\sigma_{eg}^{B''}+\mathrm{H.c.}),\label{eqa12b}\\
H^{A''B''}=&&\!\!\!\!Je^{-d/L}\frac{\Gamma_{_{1D,1}}-\Gamma_{_{1D,2}}}{\Gamma_{_{1D,1}}+\Gamma_{_{1D,2}}}(\sigma_{eg}^{A''}\sigma_{ge}^{B''}+\mathrm{H.c.}),\label{eqa12c}
\end{eqnarray}
with $\Gamma^{A''}=\Gamma_{_{1D,1}}+\Gamma_{_{1D,2}}+\Gamma'$ and $\Gamma^{B''}=\Gamma'$.
Compared with the symmetric case, here, $H^{A''B''}$ appears in the equivalent Hamiltonian, which indicates that
the asymmetry introduces a pathway between the dressed states $|A''\rangle$ and $|B''\rangle$.
We define the factor $\xi=|\Gamma_{_{1D,1}}-\Gamma_{_{1D,2}}|/(\Gamma_{_{1D,1}}+\Gamma_{_{1D,2}})$ for quantifying the asymmetry in the decay rate $\Gamma_{_{1D}}$.
The interaction strength between the dressed states $|A''\rangle$ and $|B''\rangle$ is proportional to the factor $\xi$. In Fig. \ref{figure7}(c), we give the populations of the dressed states $|A''\rangle$ and $|B''\rangle$ for three cases, when the left atom is initially excited with no driving field. The results reveal that, although the decay rate of state $|B''\rangle$ is robust to the asymmetry of the decay rates, its population also decays slowly in the asymmetric cases ($\xi\neq0$), even with $\Gamma'=0$. Intuitively, this is due to the presence of the interactions between $|A''\rangle$ and $|B''\rangle$, which can be verified by the oscillations in the population of state $|A''\rangle$, as shown in the inset of Fig. \ref{figure7}(c).
In experiment, the maximum value of the factor $\xi$ is $0.14$ for the waveguide QED superconducting circuit \cite{zhang2023sci,Mirhos2019}.
In fact, the effect of asymmetry in the anti-Bragg case can be calculated using a similar method.

\section{Discussion and summary}   \label{discussion}

The quantum phenomena in our system mentioned above may be feasible in recent experiments \cite{YLiu2017,NMSun2019}. In detail, the atom-PCW system can be realized by coupling superconducting transmon qubits to coplanar waveguides with periodically alternating sections of low and high impedance. For example, in the experiment reported by Sundaresan \emph{et al.} \cite{NMSun2019}, their device contains 14 unit cells and each cell consists of two coplanar waveguide sections with different lengths ($d_{1}=1.2$ mm, $d_{2}=7.8$ mm) and impedances (25 , 124 $\Omega$), acting as a superconducting photonic crystal.
The band-edge dispersion is quadratic $\omega_{k}=\omega_{0}+\alpha d^{2}(k-k_{0})^{2}$, where $d=d_{1}+d_{2}$ and $\alpha$ denotes the band curvature at the band edge.
They placed two transmon qubits in the high-impedance sections for maximal coupling to the band edge at $\omega_{0}/2\pi=7.8$ GHz.
Each transmon qubit is individually controllable in frequency through a local flux bias line, and the bare qubit resonance frequency is $\omega_{a}/2\pi=7.73$ GHz.
Here, the characteristic length can be written as $L=d\sqrt{\alpha/\delta}$, which can be tuned by changing the band curvature $\alpha$ and the detuning parameter $\delta$.
The characteristic strength $J$ of the long-range interactions is completely
determined by the overlap of the photonic component of the bound state for one atom with the other atom, which is controllable via the atomic frequencies.
In detail, when atomic frequencies are closer to the band edge, the bound states become more extended, enlarging the overlap and thus enhancing the long-range interactions.
The ability to tune long-range coherent interactions between atoms via the bound states shows the versatility of the platform.
Up to now, we set $\Gamma'=0$ in our discussion; i.e., we neglected the dissipation of the two atoms into free space. Since the coupling between qubits and waveguide modes is strong in superconducting waveguide QED \cite{zhang2023sci}, the assumption $\Gamma'=0$ in our simulations is reasonable. In fact, our system can also be experimentally realized by an alligator PCW \cite{AGobannatc2014,JDHood2016PNAS,SPYuapl2014} or a metamaterial waveguide
formed from an array of lumped-element $LC$ microwave resonators \cite{MMirho2018,zhang2023sci}.

In conclusion, we studied the optical properties of two atoms coupled to
a band edge of a PCW. By diagonalizing the effective Hamiltonian of the system, we presented the energy levels and their decay rates in the dressed basis. In our system, subradiant and superradiant states can be produced, and quantum interference between them mediates the photon transport in the PCW. Particularly, perfect transmission with a $\pi$ phase shift may occur on resonance for the anti-Bragg case, which can be used to realize a photon-atom phase gate. Moreover, quantum beats occur in the photon-photon correlation for the anti-Bragg case, which can be controlled due to the tunability of the bound states.
Also, directional photon emission can be observed in the anti-Bragg case in our system.
In experiment, our model is feasible for superconducting microwave transmission lines, alligator PCWs or the metamaterial waveguides.

\section*{ACKNOWLEDGMENTS}

We thank W. Nie for a critical reading and H. Zhang and M. Hua for stimulating discussions. This work is supported by the Tianjin Natural Science Foundation under Grant No. 23JCQNJC00560 and the National Natural Science Foundation of China under Grants No. 12004281 and No. 62371038.

\appendix
\setcounter{equation}{0}
\renewcommand{\theequation}{A\arabic{equation}}
\section*{Appendix}

In our system, the dynamics of the light field in a 1D waveguide is described by the Hamiltonian
\begin{equation}
H_{w}=iv_{g} \int dx \left[ a_{l}^\dagger(x) \frac{\partial a_{l}(x)}{\partial x} - a_{r}^\dagger(x) \frac{\partial a_{r}(x)}{\partial x} \right],
\end{equation}
where $v_{g}$ is the group velocity of the field and $a_{l}$ $(a_{r})$ denotes the annihilation operator of the left (right) propagating field.
The free energy of the atom dimer can be described by $H_a = \sum_{j=1}^{2} \omega_a \sigma_j^+ \sigma_j^-$, where $\omega_{a}$ is the atomic resonance frequency. The interaction between the atom dimer and the waveguide modes is described by
\begin{equation}
H_{int}\!=\!-\tilde{g}\!\int\!dx \sum_{j=1}^{2} \delta(x\!-\!x_j) \left\{ \sigma_{eg}^j \left[ a_{r}(x)\!+\! a_{l}(x) \right] \!+\!\mathrm{ H.c. } \right\},
\end{equation}
where $x_{j}$ represents the position of atom $j$ and the coupling constant $\tilde{g} = \sqrt{2\pi} g$ is assumed to be identical for all modes, with $g$ being the coupling strength between the atom and the waveguide modes. Besides the decay into the waveguide modes, decay channels to free space may exist, and the corresponding Hamiltonian is $H_{f}$.

Specifically, when excited atoms are trapped near a PCW and their resonance frequencies $\omega_{a}$ lie in the band gap,
they cannot emit photons into the PCW and may generate exponentially decaying cavity modes, which can cause long-range coherent interactions between the atoms. The interaction between the cavity modes and the atom dimer can be described by
\begin{equation}
H_I (t) = \sum_{j=1}^{2} \int dk \, g_k \sigma_{eg}^j {a}_k E_{k}(x_{j}) e^{-i\Delta t} + \mathrm{H.c.}  ,
\end{equation}
where $g_k$ denotes the coupling strength between the atom and the cavity mode with wave vector $k$, and $\Delta=\omega_{p}-\omega_{a}$ is the detuning between the frequency $\omega_{p}$ of the probe field and the atomic resonance frequency $\omega_{a}$. Due to the periodicity of photonic crystals, the guided mode can be described in Bloch's form as $E_{k}(x_{j})=e^{ikx_{j}}u_{k}(x_{j})$, where $u_{k}$ is a function with periodicity given by the lattice constant \cite{JSDouglas2015}.

Now we focus on the photon scattering of the atom dimer in a PCW. In the interaction picture, the Hamiltonian of the whole system can be expressed as $H(t) =  H'_{int}(t) + H_{I}(t) + H_{f}$ with $H'_{int}(t) \; = \; e^{iH_{w}t} H_{int} e^{-iH_{w}t}$. We assume that the input (output) state of the waveguide system and the environment is $|\psi_{in(out)}\rangle=|\varphi_{in(out)}\rangle_{w}|0\rangle_{en}$, where $|0\rangle_{en}$ is the vacuum state of the environment. Considering the single-photon scattering, the input and output states of the waveguide system are $|\varphi_{in}\rangle_w=e^{i \sigma_{\alpha} \omega_{p} t_{i}}|1_{p \alpha}\rangle$ and
$|\varphi_{out}\rangle_w=e^{i \sigma_{\beta} \omega_{p} t_{f}}|1_{p \beta}\rangle$, respectively, where $\alpha,\beta=l,r$ label the propagation directions of
the input and output photons with $\sigma_{r}=1$ and $\sigma_{l}=-1$. Here, $t_{i}$ and $t_{f}$ are the initial and final times of the scattering process with $T=t_{f}-t_{i}$, and $|1_{p \alpha (\beta)}\rangle$ represents the single-photon state of the waveguide corresponding to the propagation direction $\alpha (\beta)$.
Thus, the transition amplitude between the input and output states can be expressed as
\begin{equation}
\mathcal{A}(T) = \langle G | \langle \psi_{\text{out}} | O(T) | \psi_{\text{in}} \rangle | G \rangle,
\end{equation}
with the time evolution operator $O(T) = \mathcal{T} \exp \left[ -i \int_{t_i}^{t_i+T} dt H(t) \right]$, where $\mathcal{T}$ is the time-ordering operator. The single-photon state of the waveguide is represented by an unnormalized coherent state $|N_{p\alpha(\beta)}\rangle = \sum_{n_{p}} N_{p \alpha(\beta)}^{n_{p}} |n_{p \alpha(\beta)}\rangle / \sqrt{n_{p \alpha(\beta)}!}$, and we obtain
\begin{align}
|1_{p \alpha(\beta)}\rangle &= \lim_{N_{p \alpha(\beta)} \to 0} \frac{\delta}{\delta N_{p \alpha(\beta)}} |\{N_{p \alpha(\beta)}\}\rangle,
\end{align}
where $\{N_{p\alpha}\} = \{N_{p_1\alpha}, N_{p_2\alpha}, \ldots\}$ and $\{N_{p\beta}\} = \{N_{p_1\beta}, N_{p_2\beta}, \ldots\}$ are the number distribution of photons with different momenta $p_i$ in the initial and final states, respectively.
By introducing the displacement transformation $\tilde{O}$, $\left|\{ N_{p \alpha(\beta)} \}\right\rangle=\exp(\Sigma_{p\alpha(\beta)}|N_{p \alpha(\beta)}|^{2}/2) \tilde{O} \left| \{ 0_{p\alpha(\beta)}\} \right\rangle$, the transition amplitude reads
\begin{equation}
\mathcal{A}(T) = \lim_{N_{p \alpha(\beta)} \to 0} \left( \frac{\delta}{\delta N_{p \beta}} \right)^* \frac{\delta}{\delta N_{p \alpha}} \mathcal{A}_N(T),
\end{equation}
with
\begin{eqnarray}
\mathcal{A}_N(T) &=&\exp\left( |N_{p\alpha}|^2 e^{-i \sigma_{\alpha} \omega_{p} T} + |N_{p\beta}|^2 e^{-i \sigma_{\beta} \omega_{p} T} \right) \nonumber\\
 && \times \langle G | \langle0_{p \beta}|_{en}\langle0 | \mathcal{O}(T) | 0\rangle_{en}|0_{p \alpha} \rangle| G \rangle.
\end{eqnarray}
Note that different from the time operator $O(T)$ in Eq. (A4), the displaced time evolution operator
$\mathcal{O}(T)=\mathcal{T} \exp \big\{-i \int_{t_i}^{t_i+T} dt [ H(t) + H_d(t) ] \big\}$ contains an effective driving term
\begin{equation}
H_d (t) = -\sum_{j=1}^{2} \Omega_p (e^{ik_p x_j} \sigma_{eg}^j e^{-i\omega_{p}t} + e^{-ik_p x_j} \sigma_{ge}^j e^{i\omega_{p}t}).
\end{equation}
Then, tracing out the environment part and waveguide modes, we obtain
\begin{eqnarray}
\mathcal{A}_N(T)& = \exp\left( |N_{p\alpha}|^2 e^{-i \sigma_{\alpha} \omega_{p} T} + |N_{p\beta}|^2 e^{-i \sigma_{\beta} \omega_{p} T} \right) \nonumber\\
& \times \langle G| \mathcal{T}\exp\big\{\!\!-\!i\int_{t_i}^{t_f} dt [ H_{1} \!+\! H_d(t) ]\big\} |G\rangle,\;\;
\end{eqnarray}
with $H_{1}$ being the effective Hamiltonian in Eq. {(\ref{eqa1})}, which contains the decay to free space term described by the Hamiltonian $H'_{f}=-{{\sum\limits_{j=1}^2}}i\frac{ \Gamma'}{2} \sigma^{j}_{ee}$.
By taking the derivative of  the above equation and applying the quantum regression theorem \cite{CanevaNJP2015}, we can obtain the single-photon transmission and reflection amplitudes
\begin{align}
t &= 1 + i\frac{\Gamma_{_{1D}}}{2} \sum_{j,k=1}^{2}G_{jk} \exp\left[i\sigma_{\alpha}k_a(-x_j + x_k)\right], \\
r &= i\frac{\Gamma_{_{1D}}}{2} \sum_{j,k=1}^{2}G_{jk} \exp\left[i\sigma_{\alpha}k_a(x_j + x_k)\right],
\end{align}
where 
$G_{jk}$ are matrix elements of the Green's function $G = 1/H_{1}$, with $H_{1}$ being the effective non-Hermitian Hamiltonian in Eq. (\ref{eqa1}).
Thus, the transmission and reflection amplitudes can be written as
\begin{eqnarray}
t &=& 1 + \frac{i\Gamma_{_{1D}}}{2} \sum_j \frac{V^\dagger |\psi_j^R\rangle \langle \psi_j^L | V}{E_j}, \label{transmission}\\
r &=& \frac{i\Gamma_{_{1D}}}{2} \sum_j \frac{V^\top |\psi_j^R\rangle \langle \psi_j^L | V}{E_j}, \label{reflection}
\end{eqnarray}
respectively, where $V = \left(e^{ik_{\alpha}x_1}, e^{ik_{\alpha}x_2}\right)^\top$ denotes traveling photons in the 1D waveguide, and the right and left eigenvectors $|\psi_j^R\rangle$ and $|\psi_j^L\rangle$ of $H_{1}$ form the biorthogonal basis, i.e., $\langle \psi_j^L | \psi_{j'}^R \rangle = \delta_{jj'}$ \cite{brody}.
Specifically, in our system, the two right eigenvectors are $|\psi_{1}^{R}\rangle = (-|eg\rangle + |ge\rangle)/\sqrt{2}$ and $|\psi_{2}^{R}\rangle = (|eg\rangle + |ge\rangle)/\sqrt{2}$, and the two corresponding left vectors are obtained by transposing these right vectors. Diagonalizing the effective Hamiltonian $H_{1} = {{\sum_{j=1}^{2}}}E_{j}|\psi_{j}^{R}\rangle\langle\psi_{j}^{L}|$, we get two eigenvalues, which can be expressed as
\begin{eqnarray}
E_1 \!\!&=& \!\!-\Delta + J + J e^{-d/L} - \frac{\Gamma_{_{1D}}}{2} \sin(k_a d) - i \frac{\Gamma^A}{2}, \;\;\;\;\;\;\;\\
E_2 \!\!&=& \!\!-\Delta + J - J e^{-d/L} + \frac{\Gamma_{_{1D}}}{2} \sin(k_a d) - i \frac{\Gamma^B}{2},\;\;\;\;\;\;\;
\end{eqnarray}
where $\Gamma^A = \Gamma_{_{1D}} + \Gamma' - \Gamma_{_{1D}} \cos(k_a d)$ and $
\Gamma^B = \Gamma_{_{1D}} + \Gamma' + \Gamma_{_{1D}} \cos(k_a d)$. Under the condition $2Je^{-d/L} = \Gamma_{_{1D}}$ with $\Gamma' = 0$, by substituting the aforementioned eigenvectors and eigenvalues into Eq. (\ref{transmission}), we obtain the transmission amplitude of the incident photon in the anti-Bragg case $k_{a}d=0.5\pi$ as
\begin{eqnarray}
t=\frac{\Delta-J-i\Gamma_{_{1D}}/2}{\Delta-J+i\Gamma_{_{1D}}/2},
\end{eqnarray}
which is Eq. (\ref{eqa5plus}) in the main text.

\end{CJK*}

\begin{thebibliography}{112}





\bibitem{PLodahl2015rmp}  P. Lodahl, S. Mahmoodian, and S. Stobbe, Interfacing single photons and single quantum dots with photonic nanostructures, Rev. Mod. Phys. \textbf{87}, 347 (2015).



\bibitem{Haroche2006}   S. Haroche and J. M. Raimond, \emph{Exploring the Quantum: Atoms, Cavities, and Photons} (Oxford University Press, Oxford, 2006).



\bibitem{2006Walther}    H. Walther, B. T. H. Varcoe, B. G. Englert, and T. Becker, Cavity quantum electrodynamics,  Rep. Prog. Phys. \textbf{69}, 1325 (2006).




\bibitem{Reise2015}   A. Reiserer and G. Rempe, Cavity-based quantum networks with single atoms and optical photons, Rev. Mod. Phys. \textbf{87}, 1379 (2015).



\bibitem{Shen2007PRL}   J. T. Shen and S. Fan, Strongly correlated two-photon transport in a one-dimensional waveguide coupled to a two-level system, Phys. Rev. Lett. \textbf{98}, 153003 (2007).



\bibitem{Diby2017rmp}  D. Roy, C. M. Wilson, and O. Firstenberg, Strongly interacting photons in one-dimensional continuum, Rev. Mod. Phys. \textbf{89}, 021001 (2017).


\bibitem{DEChang2018}  D. E. Chang, J. S. Douglas, A. Gonz\'{a}lez-Tudela, C.-L. Hung, and H. J. Kimble, Colloquium: Quantum matter built from nanoscopic lattices of atoms and photons, Rev. Mod. Phys. \textbf{90}, 031002 (2018).


\bibitem{Shere2023Rev}   A. S. Sheremet, M. I. Petrov, I. V. Iorsh, A. V. Poshakinskiy, and A. N. Poddubny, Waveguide quantum electrodynamics: Collective radiance and photon-photon correlations,   Rev. Mod. Phys. \textbf{95}, 015002 (2023).


\bibitem{DayanScience2008}   B. Dayan, A. S. Parkins, T. Aoki, E. P. Ostby, K. J. Vahala, and H. J. Kimble, A photon turnstile dynamically regulated by one atom, Science \textbf{319}, 1062 (2008).



\bibitem{Vetsch2010prl}  E. Vetsch, D. Reitz, G. Sagu\'{e}, R. Schmidt, S. T. Dawkins, and A. Rauschenbeutel, Optical interface created by laser-cooled atoms trapped in the evanescent field surrounding an optical nanofiber, Phys. Rev. Lett. \textbf{104}, 203603 (2010).



\bibitem{RausPRL2011}  S. T. Dawkins, R. Mitsch, D. Reitz, E. Vetsch, and A. Rauschenbeutel, Dispersive optical interface based on nanofiber-trapped atoms, Phys. Rev. Lett.  \textbf{107}, 243601 (2011).



\bibitem{DReitz2013PRL}  D. Reitz, C. Sayrin, R. Mitsch, P. Schneeweiss, and A. Rauschenbeutel, Coherence properties of nanofiber-trapped cesium atoms, Phys. Rev. Lett. \textbf{110}, 243603 (2013).


\bibitem{Petersen2014}  J. Petersen, J. Volz, and A. Rauschenbeutel, Chiral nanophotonic waveguide interface based on spin-orbit interaction of light, Science \textbf{346}, 67 (2014).



\bibitem{Liao2015pra}    Z. Liao, X. Zeng, S. Y. Zhu, and M. S. Zubairy, Single-photon transport through an atomic chain coupled to a one-dimensional nanophotonic waveguide, Phys. Rev. A \textbf{92}, 023806 (2015).



\bibitem{Liao2016}  Z. Liao, X. Zeng, H. Nha, and M. S. Zubairy, Photon transport in a one-dimensional nanophotonic waveguide QED system, Phys. Scr. \textbf{91}, 063004 (2016).



\bibitem{HLsorensen2016}  H. L. S{\o}rensen, J. B. B\'{e}guin, K. W. Kluge, I. Iakoupov, A. S. S{\o}rensen, J. H. M\"{u}ller, E. S. Polzik, and J. Appel, Coherent backscattering of Light off one-dimensional atomic strings, Phys. Rev. Lett. \textbf{117}, 133604 (2016).



\bibitem{Cheng2017pra}   M. T. Cheng, J. P. Xu, and G. S. Agarwal, Waveguide transport mediated by strong coupling with atoms, Phys. Rev. A \textbf{95}, 053807 (2017).


\bibitem{song2017pra}   G. Z. Song, E. Munro, W. Nie, F. G. Deng, G. J. Yang, and L. C. Kwek, Photon scattering by an atomic ensemble coupled to a one-dimensional nanophotonic waveguide, Phys. Rev. A \textbf{96}, 043872 (2017).


\bibitem{PSolano2017}  P. Solano, P. B. Blostein, F. K. Fatemi, L. A. Orozco, and S. L. Rolston, Super-radiance reveals infinite-range dipole interactions through a nanofiber, Nat. Commun. \textbf{8}, 1857 (2017).



\bibitem{Litao2018}   T. Li, A. Miranowicz, X. Hu, K. Xia, and F. Nori, Quantum memory and gates using a $\Lambda$-type quantum emitter coupled to a chiral waveguide, Phys. Rev. A \textbf{97}, 062318 (2018).



\bibitem{song2021OE}    G. Z. Song, J. L. Guo, W. Nie, L. C. Kwek, and G. L. Long, Optical properties of a waveguide-mediated chain of randomly positioned atoms, Opt. Express \textbf{29}, 1903 (2021).



\bibitem{Liedl2023prl}    C. Liedl, S. Pucher, F. Tebbenjohanns, P. Schneeweiss, and A. Rauschenbeutel, Collective radiation of a cascaded Quantum system: From timed Dicke states to inverted ensembles, Phys. Rev. Lett. \textbf{130}, 163602 (2023).



\bibitem{TLHansen2008}    T. Lund-Hansen, S. Stobbe, B. Julsgaard, H. Thyrrestrup, T. S\"{u}nner, M. Kamp, A. Forchel, and P. Lodahl, Experimental realization of highly efficient broadband coupling of single quantum dots to a photonic crystal waveguide, Phys. Rev. Lett. \textbf{101}, 113903 (2008).



\bibitem{Chang2011njp} D. E. Chang, A. H. Safavi-Naeini, M. Hafezi, and O. Painter, Slowing and stopping light using an  optomechanical crystal array, New J. Phys. \textbf{13}, 023003 (2011).



\bibitem{MArcari2014prl}  M.  Arcari,  I.  S\"{o}llner,  A.  Javadi,  S.  Lindskov  Hansen,  S. Mahmoodian, J. Liu, H. Thyrrestrup, E. H. Lee, J. D. Song, S.  Stobbe,  and  P.  Lodahl, Near-unity coupling efficiency of a quantum emitter to a photonic crystal waveguide, Phys.  Rev.  Lett.  \textbf{113},  093603  (2014).


\bibitem{AGoban2015PRL}  A. Goban, C.-L. Hung, J. D. Hood, S.-P. Yu, J. A. Muniz, O. Painter, and H. J. Kimble, Superradiance for atoms trapped along a photonic crystal waveguide, Phys. Rev. Lett. \textbf{115}, 063601 (2015).



\bibitem{TudelaNAT2015}   A. Gonz\'{a}lez-Tudela, C. L. Hung, D. E. Chang, J. I. Cirac, and H. J. Kimble, Subwavelength vacuum lattices and atom-atom interactions in two-dimensional photonic crystals, Nat. Photonics \textbf{9}, 320 (2015).


\bibitem{Yu2019pnas}   S. P. Yu, J. A. Muniz, C. L. Hung, and H. J. Kimble, Two-dimensional photonic crystals for engineering atom-light interactions, Proc. Natl. Acad. Sci. U.S.A. \textbf{116}, 12743 (2019).




\bibitem{Burgersa2019}  A. P. Burgersa, L. S. Penga, J. A. Muniza, A. C. McClunga, M. J. Martina, and H. J. Kimble, Clocked atom delivery to a photonic crystal waveguide, Proc. Natl. Acad. Sci. U.S.A. \textbf{116}, 456 (2019).



\bibitem{Lzhou2008}  L. Zhou, Z. R. Gong, Y. X. Liu, C. P. Sun, and F. Nori, Controllable scattering of a single photon inside a one-dimensional resonator waveguide, Phys. Rev. Lett. \textbf{101}, 100501 (2008).


\bibitem{Liao2010pra}  J. Q. Liao, Z. R. Gong, L. Zhou, Y. X. Liu, C. P. Sun, and F. Nori, Controlling the transport of single photons by tuning the frequency of either one or two cavities in an array of coupled cavities, Phys. Rev. A \textbf{81}, 042304
(2010).



\bibitem{Rabl2016pra}  G. Calaj\'{o}, F. Ciccarello, D. E. Chang, and P. Rabl, Atom-field dressed states in slow-light waveguide QED, Phys. Rev. A \textbf{93}, 033833 (2016).



\bibitem{wang2020PRL}  Z. Wang, T. Jaako, P. Kirton, and P. Rabl, Supercorrelated radiance in nonlinear photonic waveguides, Phys. Rev. Lett. \textbf{124}, 213601 (2020).


\bibitem{Yang2020OE}  Y. H. Zhou, X. Y. Zhang, D. D. Zou, Q. C. Wu, B. L. Ye, Y. L. Fang, H. Z. Shen, and C. P. Yang, Controllable scattering of a single photon inside a one-dimensional coupled resonator waveguide with second-order nonlinearity, Opt. Express \textbf{28}, 1249 (2020).




\bibitem{Wang2023PRA}  X. J. Zhang, C. G. Liu, Z. R. Gong, and Z. H. Wang, Quantum interference and controllable magic cavity QED via a giant atom in coupled resonator waveguide, Phys. Rev. A \textbf{108}, 013704 (2023).



\bibitem{Chang2007nap}  D. E. Chang, A. S. S{\o}rensen, E. A. Demler, and M. D. Lukin, A single-photon transistor using nanoscale surface plasmons, Nat. Phys. \textbf{3}, 807 (2007).


\bibitem{AkimovNature2007}  A. V. Akimov, A. Mukherjee, C. L. Yu, D. E. Chang, A. S. Zibrov, P. R. Hemmer, H. Park, and M. D. Lukin, Quantum matter built from nanoscopic lattices of atoms and photons, Nature (London) \textbf{450}, 402 (2007).



\bibitem{Tudela2011prl}   A. Gonzalez-Tudela, D. Martin-Cano, E. Moreno, L. Martin-Moreno, C. Tejedor, and F. J. Garcia-Vidal, Entanglement of two qubits mediated by one-dimensional plasmonic waveguides, Phys. Rev. Lett. \textbf{106}, 020501 (2011).



\bibitem{Akselrod2014}   G. M. Akselrod, C. Argyropoulos, T. B. Hoang, C. Cirac\`{\i}, C.
Fang, J. Huang, D. R. Smith, and M. H. Mikkelsen, Probing the mechanisms of large Purcell enhancement in plasmonic nanoantennas, Nat. Photonics \textbf{8}, 835 (2014).



\bibitem{Anj2024PRA}    C. J. Yang, X. Y. Liu, S. Q. Xia, S. Y. Bai, and J. H. An,  Non-Markovian quantum interconnect formed by a surface plasmon polariton waveguide,
Phys. Rev. A \textbf{109}, 033518 (2024).


\bibitem{WallraffNature2004}   A. Wallraff, D. I. Schuster, A. Blais, L. Frunzio, R. S. Huang, J. Majer, S. Kumar,
S. M. Girvin, and R. J. Schoelkopf, Strong coupling of a single photon to a superconducting qubit using circuit quantum electrodynamics, Nature (London) \textbf{431}, 162 (2004).


\bibitem{ShenPRL2005}  J. T. Shen and S. Fan, Coherent single photon transport in a one-dimensional waveguide coupled with superconducting quantum bits, Phys. Rev. Lett. \textbf{95}, 213001 (2005).






\bibitem{AstafievScience2010}  O. Astafiev, A. M. Zagoskin, A. A. Abdumalikov,J r., Y. A. Pashkin,
T. Yamamoto, K. Inomata, Y. Nakamura, and J. S. Tsai, Resonance fluorescence of a single artificial atom, Science \textbf{327}, 840 (2010).


\bibitem{LooSci2013}   A. F. van Loo, A. Fedorov, K. Lalumi\'{e}re, B. C. Sanders, A. Blais, and A. Wallraff, Photon-mediated interactions between distant artificial atoms, Science \textbf{342}, 1494 (2013).


\bibitem{GuPR2017}   X. Gu, A. F. Kockum, A. Miranowicz, Y. X. Liu, and F. Nori, Microwave photonics with superconducting quantum circuits, Phys. Rep. \textbf{718-719}, 1 (2017).



\bibitem{Das2017PRL}    S. Das, V. E. Elfving, S. Faez, and A. S. S{\o}rensen, Interfacing superconducting qubits and single optical photons using molecules in waveguides, Phys. Rev. Lett. \textbf{118}, 140501 (2017).



\bibitem{Kockum2018}   A. F. Kockum, G. Johansson, and F. Nori, Decoherence-free interaction between giant atoms in waveguide quantum electrodynamics, Phys. Rev. Lett. \textbf{120}, 140404 (2018).


\bibitem{Liaoyin2022pra}    X. L. Yin, W. B. Luo, and J. Q. Liao, Non-markovian disentanglement dynamics in double-giant-atom waveguide-QED systems, Phys. Rev. A \textbf{106}, 063703 (2022).



\bibitem{Liao2023pra}    X. L. Yin and J. Q. Liao, Generation of two-giant-atom entanglement in waveguide-QED systems, Phys. Rev. A \textbf{108}, 023728  (2023).




\bibitem{Joshi2023prx}     C. Joshi, F. Yang, and M. Mirhosseini, Resonance fluorescence of a chiral artificial atom, Phys. Rev. X \textbf{13}, 021039 (2023).


\bibitem{Kannan2023NP}    B. Kannan, A. Almanakly, Y. Sung, A. Di Paolo, D. A. Rower, J. Braum\"{u}ller, A. Melville, B. M. Niedzielski, A. Karamlou, K. Serniak, A. Veps\"{a}l\"{a}inen, M. E. Schwartz, J. L. Yoder, R. Winik, J. I. J. Wang, T. P. Orlando, S. Gustavsson, J. A. Grover, and W. D. Oliver, On-demand directional microwave photon emission using waveguide quantum electrodynamics, Nat. Phys. \textbf{19}, 394 (2023).



\bibitem{Jingj2024NJP}     J. Li, J. Lu, Z. R. Gong and L. Zhou, Tunable chiral bound states in a dimer chain of coupled resonators, New J. Phys. \textbf{26}, 033025 (2024).



\bibitem{klalum2013}   K. Lalumi\`{e}re, B. C. Sanders, A. F. van Loo, A. Fedorov, A. Wallraff, and A. Blais, Input-output theory for waveguide QED with an ensemble of inhomogeneous atoms, Phys. Rev. A \textbf{88}, 043806 (2013).


\bibitem{zhang2019prl}    Y. X. Zhang and K. M{\o}lmer, Theory of subradiant states of a one-dimensional two-level atom chain, Phys. Rev. Lett. \textbf{122}, 203605 (2019).


\bibitem{Henriet2019pra} L. Henriet, J. S. Douglas, D. E. Chang, and A. Albrecht, Critical open-system dynamics in a one-dimensional optical-lattice clock, Phys. Rev. A \textbf{99}, 023802 (2019).



\bibitem{Ke2019prl}    Y. Ke, A. V. Poshakinskiy, C. Lee, Y. S. Kivshar, and A. N. Poddubny, Inelastic scattering of photon pairs in qubit arrays with subradiant states, Phys. Rev. Lett. \textbf{123}, 253601 (2019).


\bibitem{Albrecht2019njp}    A. Albrecht, L. Henriet, A. Asenjo-Garcia, P. B. Dieterle, O. Painter, and D. E. Chang, Subradiant states of quantum bits coupled to a one-dimensional waveguide, New J. Phys. \textbf{21}, 025003 (2019).



\bibitem{zhang2020prr}   Y. X. Zhang, C. Yu, and K. M{\o}lmer, Subradiant bound dimer excited states of emitter chains coupled to a one dimensional waveguide, Phys. Rev. Res. \textbf{2}, 013173 (2020).



\bibitem{Nie2020prl}   W. Nie, Z. H. Peng, F. Nori, and Y. X. Liu, Topologically protected quantum coherence in a superatom, Phys. Rev. Lett. \textbf{124}, 023603 (2020).


\bibitem{Holzin2022prl}    R. Holzinger, R. Guti\'{e}rrez-J\'{a}uregui, T. H\"{o}nigl-Decrinis, G. Kirchmair, A. Asenjo-Garcia, and H. Ritsch, Control of localized single- and many-body dark states in waveguide QED, Phys. Rev. Lett. \textbf{129}, 253601 (2022).


\bibitem{Tiranov2023sci}   A. Tiranov, V. Angelopoulou, C. J. van Diepen, B. Schrinski, O. A. D. Sandberg, Y. Wang, L. Midolo, S. Scholz, A. D. Wieck, A. Ludwig, A. S. S{\o}rensen, and P. Lodahl, Collective super- and subradiant dynamics between distant optical quantum emitters, Science \textbf{379}, 389 (2023).



\bibitem{Nie2023PRL}    W. Nie, T. Shi, Y. X. Liu, and F. Nori, Non-Hermitian waveguide cavity QED with tunable atomic mirrors, Phys. Rev. Lett. \textbf{131}, 103602 (2023).




\bibitem{Nie2024ctp}  H. L. Cheng and W. Nie. Collectively induced transparency and absorption in waveguide QED with Bragg atom arrays, Commun. Theor. Phys. \textbf{76}, 085101 (2024).


\bibitem{Lu2024qst}   Y. W. Lu, J. F. Liu, H. Jiang, and Z. Liao, Topologically protected subradiant cavity polaritons through linewidth narrowing
enabled by dissipationless edge states, Quantum Sci. Technol. \textbf{9}, 035019 (2024).


\bibitem{Lu2024PRA}   Y. W. Lu, Topology-empowered decoherence suppression of strong light-matter interaction in a one-dimensional atomic cavity, Phys. Rev. A \textbf{110}, 033720 (2024).









\bibitem{SJohn1990prl}  S. John and J. Wang, Quantum electrodynamics near a photonic band gap: Photon bound states and  dressed atoms, Phys. Rev. Lett. \textbf{64}, 2418 (1990).



\bibitem{SJohn1990prb}   S. John and J. Wang, Quantum optics of localized light in a photonic band gap, Phys. Rev. B \textbf{43}, 12772 (1991).


\bibitem{SJohn1995prl}   S. John and T. Quang, Localization of superradiance near a photonic band gap, Phys. Rev. Lett. \textbf{74}, 3419 (1995).






\bibitem{Hung2013}   C. L. Hung, S. M. Meenehan, D. E. Chang, O. Painter, and H. J. Kimble, Trapped atoms in one-dimensional photonic crystals,
New J. Phys. \textbf{15}, 083026 (2013).



\bibitem{ShiT2016}  T. Shi, Y.-H. Wu, A. Gonz\'{a}lez-Tudela, and J. I. Cirac, Bound states in boson impurity models, Phys. Rev. X \textbf{6}, 021027 (2016).






\bibitem{JSDouglas2015}  J. S. Douglas, H. Habibian, C. L. Hung, A. V. Gorshkov, H. J. Kimble, and  D. E. Chang, Quantum many-body models with cold atoms coupled to photonic crystals, Nat. Photonics \textbf{9}, 326 (2015).


\bibitem{jsDoug2016prx}  J. S. Douglas, T. Caneva, and D. E. Chang, Photon molecules in atomic gases trapped near photonic crystal waveguides, Phys. Rev. X \textbf{6}, 031017 (2016).


\bibitem{Shah2016Op}   E. Shahmoon, P. Gri\v{s}ins, H. P. Stimming, I. Mazets, and G. Kurizki, Highly nonlocal optical nonlinearities in atoms trapped near a waveguide, Optica \textbf{3}, 725 (2016).




\bibitem{Shi2018njp}    T. Shi, Y. H. Wu, A. Gonz\'{a}lez-Tudela, and J. I. Cirac, Effective many-body Hamiltonians of qubit-photon bound states, New J. Phys. \textbf{20}, 105005 (2018).


\bibitem{Kumar2023}    N. P. Kumar, A. R. Hamann, R. Navarathna, M. Zanner, M. Pletyukhov, and A. Fedorov,
Qubit-photon bound states: Crossover from waveguide to cavity regime, Phys. Rev. Applied \textbf{20}, 024058 (2023).



\bibitem{Song2018}  G. Z. Song, E. Munro, W. Nie, L. C. Kwek, F. G. Deng, and G. L. Long, Photon transport mediated by an atomic chain trapped along a photonic crystal waveguide, Phys. Rev. A \textbf{98}, 023814 (2018).



\bibitem{Bello2019}  M. Bello, G. Platero, J. I. Cirac, and A. Gonz\'{a}lez-Tudela, Unconventional quantum optics in topological waveguide QED, Sci. Adv. \textbf{5}, eaaw0297 (2019).



\bibitem{Song2019}   G. Z. Song, L. C. Kwek, F. G. Deng, and G. L. Long, Microwave transmission through an artificial atomic chain coupled to a superconducting photonic crystal, Phys. Rev. A \textbf{99}, 043830 (2019).


\bibitem{Kim2021prx}    E. Kim, X. Zhang, V. S. Ferreira, J. Banker, J. K. Iverson, A. Sipahigil, M. Bello, A. Gonz\'{a}lez-Tudela, M. Mirhosseini, and
O. Painter, Quantum electrodynamics in a topological waveguide, Phys. Rev. X \textbf{11}, 011015 (2021).



\bibitem{Bello2022prx}   M. Bello, G. Platero, and A. Gonz\'{a}lez-Tudela, Spin many-body phases in standard- and topological-waveguide QED simulators, PRX Quantum \textbf{3}, 010336 (2022).


\bibitem{Wang2024res}   X. Wang, H. B. Zhu, T. Liu, and F. Nori, Realizing quantum optics in structured environments with giant atoms, Phys. Rev. Research \textbf{6}, 013279 (2024).



\bibitem{YLiu2017}  Y. Liu and A. A. Houck, Quantum electrodynamics near a photonic bandgap, Nat. Phys. \textbf{13}, 48 (2017).



\bibitem{NMSun2019}  N. M. Sundaresan, R. Lundgren, G. Zhu, A. V. Gorshkov, and A. A. Houck, Interacting Qubit-Photon Bound States with Superconducting Circuits, Phys. Rev. X \textbf{9}, 011021 (2019).


\bibitem{Scigliu2022}     M. Scigliuzzo, G. Calaj\`{o}, F. Ciccarello, D. Perez Lozano, A. Bengtsson, P. Scarlino, A.Wallraff, D. Chang, P. Delsing, and S. Gasparinetti, Controlling atom-photon bound states in an array of Josephson-junction resonators, Phys. Rev. X \textbf{12}, 031036 (2022).


\bibitem{zhang2023sci}    X. Zhang, E. Kim, D. K. Mark, S. Choi, and O. Painter, A superconducting quantum simulator based on a photonic-bandgap metamaterial, Science \textbf{379}, 278 (2023).




\bibitem{ShenOL2005} J. T. Shen and S. Fan, Coherent photon transport from spontaneous emission in one-dimensional waveguides, Opt. Lett. \textbf{30}, 2001 (2005).


\bibitem{zheng2013PRL}  H. Zheng and H. U. Baranger, Persistent quantum beats and long-distance entanglement from
waveguide-mediated interactions,  Phys. Rev. Lett. \textbf{110}, 113601 (2013).



\bibitem{Zheng2010PRA}    H. Zheng, D. J. Gauthier, and H. U. Baranger, Waveguide QED: Many-body bound-state effects in coherent
and Fock-state scattering from a two-level system, Phys. Rev. A \textbf{82}, 063816 (2010).



\bibitem{JDJoan2008book}  J. D. Joannopoulos, S. G. Johnson, J. N. Winn, and R. D. Meade, \emph{Photonic Crystals: Molding the Flow of Light},
2nd ed. (Princeton University Press, Princeton, NJ, 2008).



\bibitem{EWAN2017}  E. Munro, L. C. Kwek, and D. E. Chang, Optical properties of an atomic ensemble coupled to a band edge of a photonic crystal waveguide, New J. Phys. \textbf{19}, 083018 (2017).


\bibitem{Albrecht2017njp}  A. Albrecht, T. Caneva, and D. E. Chang, Changing optical band structure with single photons, New J. Phys. \textbf{19}, 115002 (2017).


\bibitem{Chang2012}   D. E. Chang, L. Jiang, A. V. Gorshkov, and H. J. Kimble, Cavity QED with atomic mirrors, New J. Phys. \textbf{14}, 063003 (2012).


\bibitem{CanevaNJP2015}   T. Caneva, M. T. Manzoni, T. Shi, J. S. Douglas, J. I. Cirac, and D. E. Chang, Quantum dynamics of propagating photons with strong interactions: A generalized input-output formalism, New J. Phys. \textbf{17}, 113001 (2015).



\bibitem{Shitao2015}   T. Shi, D. E. Chang, and J. I. Cirac, Multiphoton-scattering theory and generalized master equations, Phys. Rev. A \textbf{92}, 053834 (2015).


\bibitem{Nie2021praapp}   W. Nie, T. Shi, F. Nori, and Y. X. Liu, Topology-enhanced nonreciprocal scattering and photon absorption in a waveguide, Phys. Rev. Applied \textbf{15}, 044041 (2021).


\bibitem{Li2012prl}    Y. Li, L. Aolita, D. E. Chang, and L. C. Kwek, Robust-fidelity atom-photon entangling gates in the weak-coupling regime, Phys. Rev. Lett. \textbf{109}, 160504 (2012).


\bibitem{Loudon2003}    R. Loudon, \emph{The Quantum Theory of Light}, 3rd ed. (Oxford University Press, New York, 1997).



\bibitem{Sanche2020prl}   C. S\'{a}nchez Mu\~{n}oz and F. Schlawin, Photon correlation spectroscopy as a witness for quantum coherence, Phys. Rev. Lett.
\textbf{124}, 203601 (2020).



\bibitem{Limonov2017NP}  M. F. Limonov, M. V. Rybin, A. N. Poddubny, and Y. S. Kivshar, Fano resonances in photonics, Nat. Photonics \textbf{11}, 543 (2017).



\bibitem{Kannan2020na}     B. Kannan, M. J. Ruckriegel, D. L. Campbell, A. Frisk Kockum, J. Braum\"{u}ller, D. K. Kim, M. Kjaergaard, P. Krantz,
A. Melville, B. M. Niedzielski, A. Veps\"{a}l\"{a}inen, R. Winik, J. Yoder, F. Nori, T. P. Orlando, S. Gustavsson, and W. D.
Oliver, Waveguide quantum electrodynamics with superconducting artificial giant atoms, Nature (London) \textbf{583}, 775 (2020).


\bibitem{Mirhos2019}    M. Mirhosseini, E. Kim, X. Zhang, A. Sipahigil, P. B. Dieterle, A. J. Keller, A. Asenjo-Garcia, D. E. Chang, and
O. Painter, Cavity quantum electrodynamics with atom-like mirrors, Nature (London) \textbf{569}, 692 (2019).



\bibitem{JDHood2016PNAS}  J. D. Hood, A. Goban, A. Asenjo-Garcia, M. Lu, S.-P. Yu, D. E. Chang, and H. J. Kimble,
Atom-atom interactions around the band edge of a photonic crystal waveguide, Proc. Natl. Acad. Sci. U.S.A. \textbf{113}, 10507 (2016).



\bibitem{AGobannatc2014}  A. Goban, C. L. Hung, S. P. Yu, J. D. Hood, J. A. Muniz, J. H. Lee,
M. J. Martin, A. C. McClung, K. S. Choi, D. E. Chang, O. Painter, and H. J. Kimble, Atom-light interactions in photonic crystals, Nat. Commun. \textbf{5}, 3808 (2014).



\bibitem{SPYuapl2014}  S.-P. Yu, J. D. Hood, J. A. Muniz, M. J. Martin, R. Norte,
C.-L. Hung, S. M. Meenehan, J. D. Cohen, O. Painter, and H. J.  Kimble, Nanowire photonic crystal waveguides for single-atom trapping and strong light-matter interactions, Appl. Phys. Lett. \textbf{104}, 111103 (2014).




\bibitem{MMirho2018}  M. Mirhosseini, E. Kim, V. S. Ferreira, M. Kalaee, A. Sipahigil, A. J. Keller, and O. Painter, Superconducting metamaterials for waveguide quantum electrodynamics, Nat. Commun. \textbf{9}, 3706 (2018).


\bibitem{brody}    D. C. Brody, Biorthogonal quantum mechanics, J. Phys. A \textbf{47}, 035305 (2013).




\end{thebibliography}
\end{document}